\documentclass[%
 reprint,
 amsmath,amssymb,
 aps,
 prl,
]{revtex4-2}
\usepackage{graphicx}
\usepackage{dcolumn}
\usepackage{bm}
\usepackage{xcolor}
\usepackage{float}

\begin{document}

\title{Subrelativistic Alternating Phase Focusing Dielectric Laser Accelerators}

\author{Payton Broaddus,$^{1}$, Thilo Egenolf,$^{3}$ Dylan S. Black,$^{1}$, Melanie Murillo,$^{1}$ Clarisse Woodahl, $^{1}$, Yu Miao,$^{1}$,  Uwe Niedermayer,$^{3}$, Robert L. Byer,$^{2}$ Kenneth J. Leedle$^1$, Olav Solgaard$^1$ }

\affiliation{
 $^1$Department of Electrical Engineering, Stanford University, \break 350 Serra Mall, Stanford, California 94305-9505, USA \break
$^2$Department of Applied Physics, Stanford University, \break 348 Via Pueblo Mall, Stanford, California 94305-4090, USA \break
 $^3$Technische Universit\"at Darmstadt, Institut f\"ur Teilchenbeschleunigung und Elektromagnetische Felder (TEMF), \break Schlo{\ss}gartenstra{\ss}e. 8, 64289 Darmstadt, Germany
}%

\begin{abstract}
We demonstrate a silicon-based electron accelerator that uses laser optical near fields to both accelerate and confine electrons over extended distances. Two dielectric laser accelerator (DLA) designs were tested, each consisting of two arrays of silicon pillars pumped symmetrically by pulse front tilted laser beams, designed for average acceleration gradients 35 and 50 MeV/m respectively. The DLAs are designed to act as alternating phase focusing (APF) lattices, where electrons, depending on the electron-laser interaction phase, will alternate between opposing longitudinal and transverse focusing and defocusing forces. By incorporating fractional period drift sections that alter the synchronous phase between $\pm 60^\circ$ off crest, electrons captured in the designed acceleration bucket experience half the peak gradient as average gradient while also experiencing strong confinement forces that enable long interaction lengths. We demonstrate APF accelerators with interaction lengths up to 708 $\mu$m and energy gains up to $23.7 \pm 1.07$ keV FWHM, a 25\% increase from starting energy, demonstrating the ability to achieve substantial energy gains with subrelativistic DLA.

\end{abstract}

\maketitle

 Dielectric laser accelerators (DLAs) utilize recent advances in semiconductor nanofabrication, high fluence femtosecond lasers, and low emittance electron sources to produce electron accelerators with acceleration gradients 1--2 orders of magnitude higher than conventional copper rf accelerators \cite{England2014,Shiloh:22}. Key to this technology is the GV/m laser-induced damage threshold of semiconductor materials, which demonstrated acceleration gradients as high as 850 MeV/m for relativistic DLAs and 370 MeV/m for subrelativistic DLAs \cite{Leedle:15,Cesar2018}. Leveraging these high gradients over long interaction lengths to produce high energy gains at subrelativistic energies has been difficult due to confinement challenges.

Subrelativistic dual pillar DLAs have the necessary transverse lensing to confine beams within submicron apertures, producing focusing forces equivalent to quadrupole focusing gradients of $1.4 \pm 0.1  \text{MT/m}$ \cite{Black2019}. Recently, elements needed to realize fully integrated DLAs have been demonstrated for dual pillar structures in the subrelativistic regime: low energy spread attosecond bunchers, and confinement lattices \cite{GermanAPF,Niedermayer2021}. Central to these successes was the application of alternating phase focusing (APF), originally developed for ion acceleration in the 1950s~\cite{Fainberg1956,Lapostolle89} and adapted for DLAs in \cite{Niedermayer2018a}, where the Lorentz force of the accelerating laser mode itself is used for confinement instead of relying on external magnets \cite{Wangler2008}.

In \cite{GermanAPF}, the first APF confinement lattice was demonstrated in DLAs, albeit without acceleration, and in \cite{Niedermayer2021}, an APF attosecond buncher was demonstrated, producing microbunches compact enough and with low enough energy spread to be injected into an APF DLA. This Letter demonstrates the realization of an accelerating APF DLA, which coherently accelerates and confines electrons over extended distances. 

Following \cite{,Niedermayer2018a}, APF DLA lattices are designed around a $\hat{z}$ traveling “synchronous electron” in the center $y = 0$ of a symmetric dual drive mode at synchronous phase $\phi_{s}$. Once an injection energy and operating synchronous phase have been selected, the energy ramp is fixed by the structure factor $|e_{1n}|$, incident laser field amplitude $E_0$, and structure periodicity~\cite{NiedermayerJournPhysConf:17}, which follows the Wideroe condition for the first spatial mode, expressed as $\Lambda = \beta_s \lambda_0$, where $\Lambda$ is the periodicity, $\beta_s$ is the ratio of the relativistic velocity to the speed of light, and $\lambda_0$ is the central wavelength. For a $\hat{x}$ invariant dual pillar DLA powered by two counterpropagating in-phase monochromatic $\hat{z}$ polarized lasers incident from $\pm \hat{y}$, the synchronous electron will gain energy $\Delta W$ per period~\cite{Leedle:18},
\begin{equation}
    \Delta W (\phi_s) = -q |e_{1n}| E_0 \Lambda  \cos(\phi_s),
    \label{Eq:energyGain}
\end{equation}
where $q$ is the elementary charge. The Lorentz force on a synchronous electron is written as
\begin{equation}
\vec{F}(\phi_s) = \begin{bmatrix} 
F_x \\
F_y\\
Fz
\end{bmatrix} =\frac{-q c}{\beta\gamma}|e_{1n}| E_0 \begin{bmatrix} 
0 \\
\frac{1}{\gamma}\sinh(\Gamma_yy)\sin\phi_s \\
\cosh(\Gamma_yy)\cos\phi_s
\end{bmatrix}.
\end{equation}

$\gamma$ is the Lorentz factor and $\Gamma_y = ik_y = \frac{2\pi}{\lambda_0}\sqrt{\frac{1}{\beta_s^2} - 1}$, where $k_y$ is the imaginary (evanescent) wavenumber for the accelerating first spatial mode. Nonsynchronous electrons experience a Lorentz force $\vec{F}(\phi)$, where the interaction phase $\phi$ can be expressed as $\phi = \phi_s - \omega_0 \Delta t$  \cite{Leedle:18}. Here, $\Delta t = t_e - t_s$ represents the difference in arrival time between a nonsynchronous electron and the synchronous electron, denoted by $t_e$ and $t_s$, respectively, and $\omega_0 = \frac{2\pi c}{\lambda_0}$.

As seen in Fig. \ref{fig:InjectionPhase}, for synchronous phase $\phi_s \in (0,\pi)$, longitudinally focusing [($\frac{\delta F_z}{\delta \phi}) < 0$] and transversely defocusing [$F_y(y > 0) > 0$] forces act on the electrons near the synchronous electron. For synchronous phase $\phi_s \in (-\pi,0)$, the opposite occurs, longitudinal defocusing and transverse focusing forces. The synchronous phase, and thus longitudinal and transverse focusing forces, can be switched via fractional drift sections \cite{Niedermayer2018a},
\begin{equation}
    l_{fd} = (2\pi - \phi_s)\Lambda/\pi, \;\;\;
    l_{df} = (\pi - \phi_s)\Lambda/\pi,
        \label{Eq:ldf}
\end{equation}
where $l_{fd}$ switches the forces on electrons near a synchronous reference particle from transversely focusing to transversely defocusing, and $l_{df}$ from transversely defocusing to transversely focusing. By using these fractional drift sections, a properly designed APF DLA can capture electrons with the correct injection phase and transport them in an accelerating bucket over extended distances. 

\begin{figure}[t]
\includegraphics[width=8.5cm]{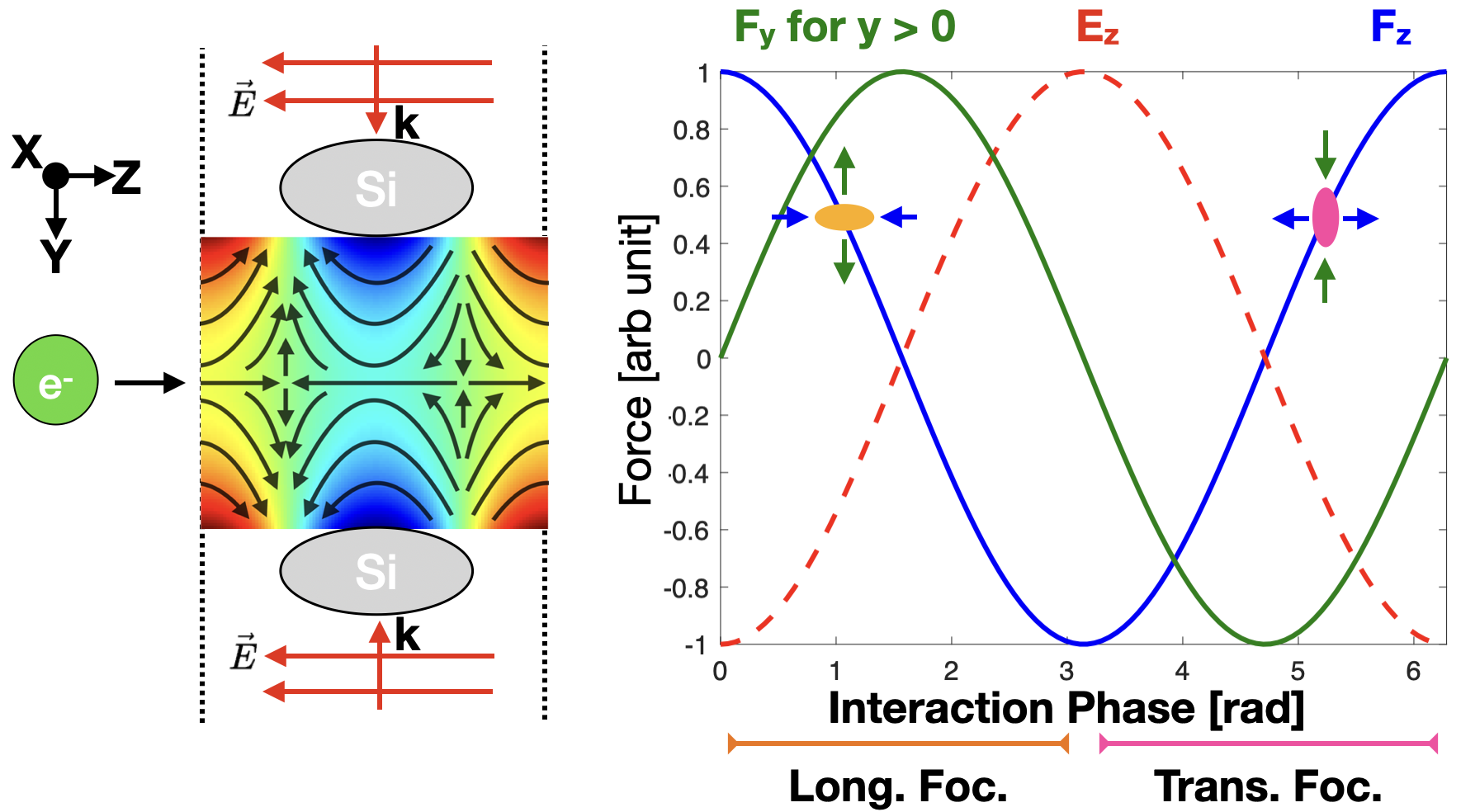}
\caption{Left: DLA period with Lorentz force vectors overlaid on electric field $-E_z$ from accelerating mode. Right: $E_z$ and Lorentz forces acting on a reference particle versus interaction phase. Two bunches are drawn at $\phi_s$ = $\pm$ $60^\circ$, each experiencing half maximum accelerating force $F_z$ and confinement forces of opposite sign in $\hat{y}$ and $\hat{z}$.}
\label{fig:InjectionPhase}
\end{figure}

Two dual pillar APF DLA lattice designs were tested, which will be referred to as DLA70 (designed at TU Darmstadt) and DLA100 (designed at Stanford). Both were designed for operation at 1980 nm, 96 keV injection energy and synchronous phase $\pm60^\circ$. Unlike \cite{GermanAPF}, which operated at synchronous phase $\pm90^\circ$ without acceleration and with full confinement force, operating at $\pm60^\circ$ provides half maximal acceleration and 86\% maximal confinement force, a good balance between maximizing both confinement and temporal acceptance and maximizing average gradient \cite{Niedermayer2018a}.

\begin{figure}[h]
\includegraphics[width=8cm]{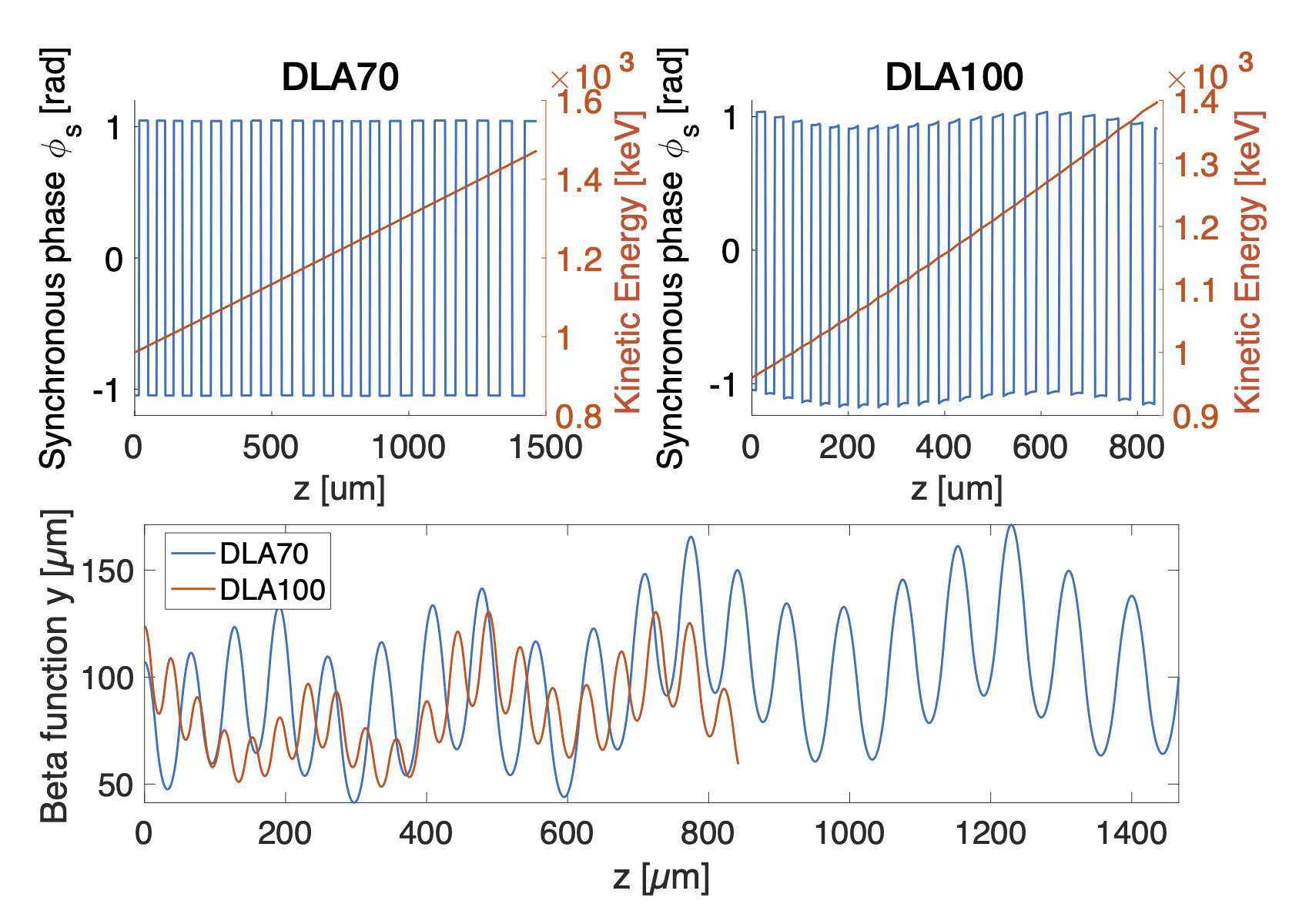}
\caption{ Synchronous phase and energy (top) and Courant-Snyder $\hat{\beta}_y$ function (bottom) for infinitesimal emittance versus travel distance extracted from DLAtrack6D simulation. The nominal electric field is 106 MV/m for DLA70 and 250 MV/m for DLA100, respectively.
}
\label{fig:APFInjection}
\end{figure}

\begin{figure}[hbt!]
\includegraphics[width=8.5cm]{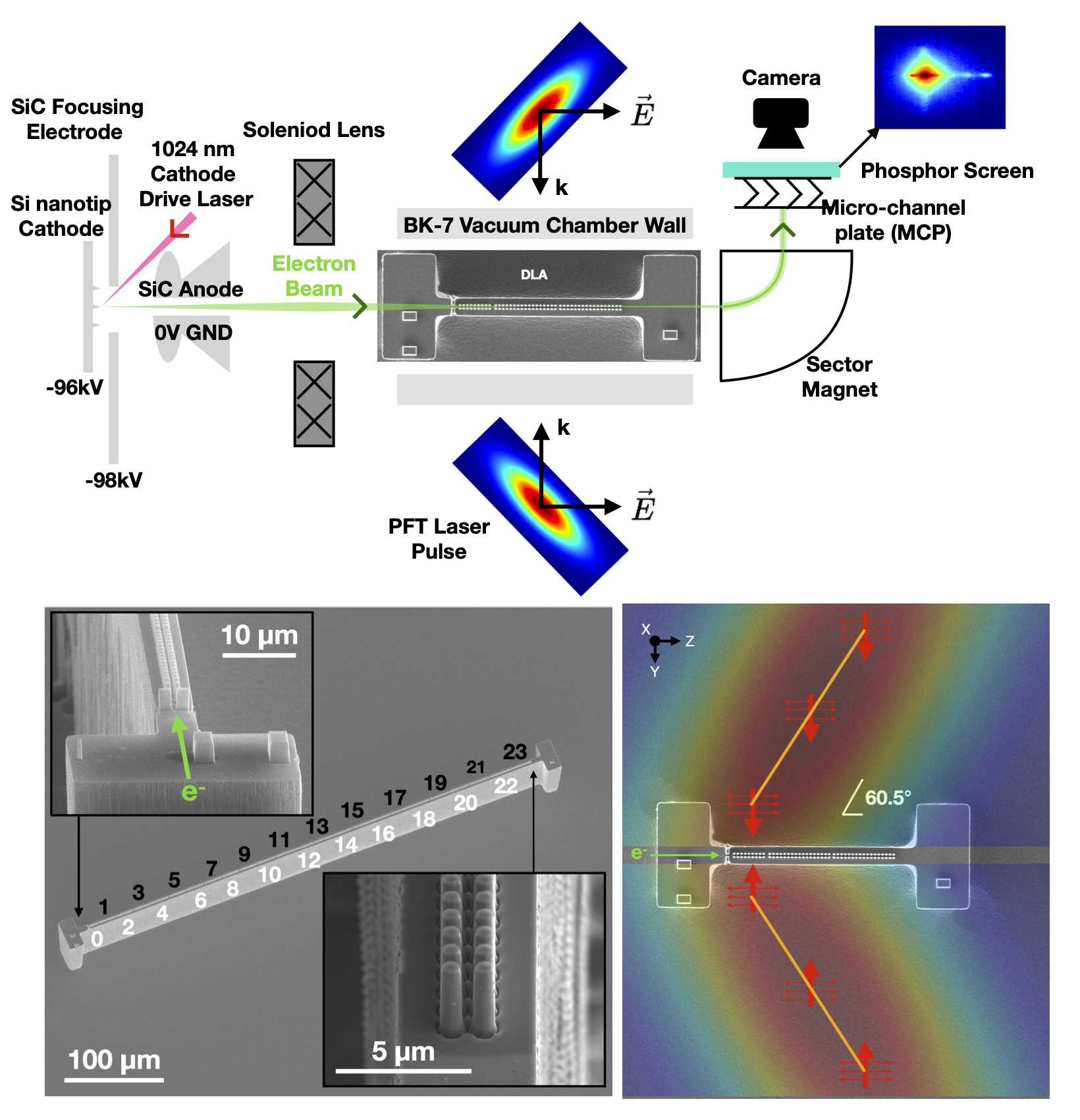}
\caption{Top: accelerator system overview. Bottom Left: DLA100 476 $\mu$m on a 500 $\mu$m long mesa. Cells labeled as black are longitudinally focusing and transversely defocusing (LFTD), while white are longitudinally defocusing and transversely focusing (LDTF). Bottom Right: DLA100 46 $\mu$m powered by two $60.5 \pm 0.7^{\circ}$ incident PFT laser pulses. The orange line shows the electron-laser overlap on the laser pulse as it travels through the structure. }
\label{fig:Overview}
\end{figure} 
\begin{figure*}[t]
\includegraphics[width=17cm]{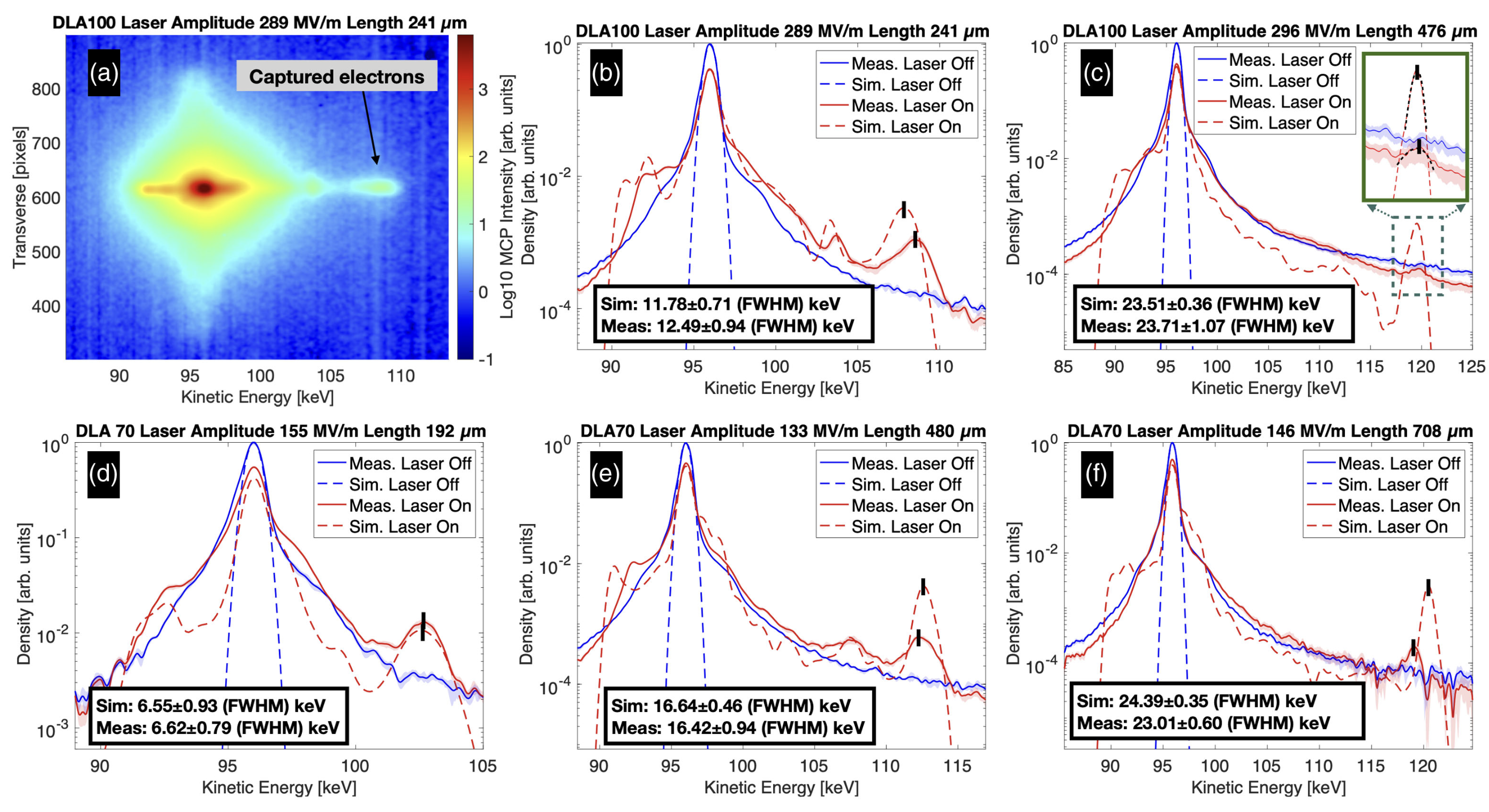}
\caption{ Simulated and measured laser-on and laser-off spectra for DLA70 and DLA100. 
(a) Integrated MCP camera image of laser on DLA70 480 $\mu$m. Plots in (b)-(f) are obtained by integrating horizontal slices near peak transverse pixel. (b),(c) Simulated and measured DLA100 spectra for laser on and laser off. (d)-(f) Simulated and measured DLA70 spectra for laser on and laser off. A semitransparent overlay on the measured spectra covers the 25\%-75\% quantile. Experimental parameters in Table \ref{tab:DLAproperties}.
}
\label{fig:MeasSim}
\end{figure*}
Each had different pillar dimensions, with structure factor $|e_{1n}|$ ranging from 0.68 to 0.78 for DLA70 and 0.37 to 0.46 for DLA100 over the length of the structure, and with optimal dual-drive in-phase peak incident electric field amplitude of 106 and 250 MV/m, respectively. DLA70 and DLA100 had initial acceleration gradients of 70 and 100 MeV/m for on-crest electrons, and 35 and 50 MeV/m for synchronous electrons. The dual pillar geometry of DLA70 exhibited high transparency, suppressing deflecting sinh modes, while DLA100's highly reflective geometry enabled the generation of deflecting sinh modes, which was observed during alignment and was consistent with prior experimental studies \cite{Leedle:18,PhysRevAccelBeams.23.114001}.

DLA70 and DLA100, with channel widths 420 and 400 nm, respectively, were optimized for minimal electron beam sidewall loss \cite{Black2022,Niedermayer2018a}. The optimization involved iterative cell length selection to minimize the Courant-Snyder $\hat{\beta}_y$ function's maximum, enabling maximum transverse 1D single particle emittance $\epsilon$ propagation. For matched beam injection, the beam waist $a(z)$ is $a(z) =\sqrt{\epsilon \hat{\beta}_y (z)}$. Both structures start with $\phi_s = -60^{\circ}$ for maximal transverse capture.

Figure \ref{fig:APFInjection} displays the energy ramps, synchronous phases, and extracted $\hat{\beta}_y$ functions for DLA70 and DLA100 driven at 1980 nm wavelength, and with DLA70 simulated at 106 MV/m and DLA100 at 250 MV/m, the amplitude with best extracted $\hat{\beta}_y$ and a slight sinusoidal phase envelope.

Figure \ref{fig:Overview} shows the experimental setup; see details on fabrication and optics in the Supplemental Material \cite{APF_DLA_Supplemental_2024}. The electron source generated 96 keV $\pm$ 23 eV FWHM, 830 $\pm$ 100 fs FWHM long electron bunches at a rep rate of 100 kHz, producing $\sim$0.5 electrons per shot with $\sim$100 pmrad transverse normalized emittance in $\hat{x}$ and $\hat{y}$ \cite{ImmersionLens2022}. Once aligned, the devices were symmetrically pumped by two laser beams, and a sector magnet was used to translate energy gain into horizontal displacement on a microchannel plate (MCP) detector \cite{leedle2016laser}.

\begin{table*}[ht]
\centering
\caption{\textbf{DLA70 and DLA100}}
\label{tab:DLAproperties}
\begin{tabular}{|p{1.16cm}|p{0.65cm}|p{1.65cm}|p{1.3cm}|p{1.85cm}|p{1.85cm}|p{2.0cm}|p{1.8cm}|p{1.9cm}|p{1.6cm}|p{1.2cm}|}
\hline
& \multicolumn{2}{c|}{\textbf{Design Params}} 
& \multicolumn{7}{c|}{\textbf{Experimental Measurements}} 
\\
\cline{2-10}
\textbf{DLA Type} 
& \textbf{Len [$\mu$m]} 
& \textbf{Initial Gradient [MeV/m]}
& \textbf{PFT $E_{max}$ [MV/m]} 
& \textbf{Wavelength [nm] (sim)} 
& \textbf{$\Delta$W[keV] (sim)} 
& \textbf{FWHM $\Delta$W[keV] (sim)} 
& \textbf{Gradient [MeV/m] (sim)}
& \textbf{Capture [ppm] (sim)}
& \textbf{Capture Std [ppm]}\\
\hline
DLA70  & 192 & 35 & 155$\pm$17 & 1980 (1980) & 6.62 (6.55) & 0.79 (0.93)& 34.5 (34.1) & 10300 (9620) & 5200\\
DLA70 & 480 & 35 & 133$\pm$14 & 1980 (1980) & 16.42 (16.64) &  0.94 (0.46) & 34.2 (34.7) & 449 (1650) & 183\\
DLA70 & 708 & 35 & 146$\pm$16 & 1978 (1980) & 23.01 (24.39) & 0.6 (0.35)& 32.5 (34.4) & 76 (658) & 61\\
\hline
DLA100 & 241 & 50 & 289$\pm$39 & 1990 (1980) & 12.49 (11.76) & 0.94 (0.70) & 51.8 (48.7)& 959 (2315) & 445\\
DLA100  & 476 & 50 & 296$\pm$40 & 1995 (1980) & 23.71 (23.51) & 1.07 (0.36) & 49.8 (49.3) & 98 (272) & 45\\
\hline
\end{tabular}
\end{table*}

To achieve full electron-laser interaction with the longest 1469 $\mu$m DLA structure, our 100 kHz, 310 fs FWHM (field) optical parametrically amplified laser pulses would have needed to be stretched to 8.4 ps using temporal pulse stretching, assuming a temporal flattop pulse. Laser-induced damage threshold scales inversely to the square root of laser pulse duration in near-infrared (NIR) picosecond laser regime, and similarly unannealed silicon pillars have been destroyed with 310 fs pulsed fields as low as 419 $\pm$ 42 MV/m  \cite{Soong2014}\cite{Miao2020}. As detailed in the Supplemental Material \cite{APF_DLA_Supplemental_2024}, we employ pulse front tilted (PFT) beams in order to keep the local pulse length short while enabling interaction in long structures at nearly constant amplitude \cite{Cesar:18}\cite{Wei:17}\cite{CESAR2018252}. The PFT angle is matched for a $\beta = 0.564$ beam, the average velocity assuming a 24 keV energy gain. Because of subrelativistic energy-velocity scaling, this approach has a maximum field amplitude error of $5\%$ for the longest structures~\cite{3DAPF}. Simulations include this effect as a uniform $5\%$ phase and amplitude error.

Figure \ref{fig:MeasSim} shows the measured and simulated (DLAtrack6D \cite{NiedermayerPRAB:17,Niedermayer:ICAP2018-MOPLG01}) MCP spectra for DLA70 and DLA100 of different lengths and peak electric fields for both laser-on and laser-off conditions, with extracted parameters shown in Table \ref{tab:DLAproperties}. Laser-on (1980 nm, amplitude as measured in the experiments) and laser-off simulations were performed and simulated, and measured spectra were normalized to their maximum laser-off signal.  The simulated and measured spectra [Figs. \ref{fig:MeasSim} (b)-(f)] all show similar features: laser-on peak depletion at 96 keV, an asymmetric shoulder modulation, and a captured electron peak, denoted by a black dash marker, with some discrepancy in total count for each feature. For DLA100 241 $\mu$m and DLA70 480 $\mu$m, clear subpeaks are also visible at 103.2 keV and at 107.3 keV, respectively.  

 The electron pulse duration (830 fs FWHM) being longer than the laser pulse duration (330 fs FWHM) results in only a portion of the electron pulse interacting and thus experiencing energy modulation from the laser-on signal. This causes peak depletion for the noninteracting injection energy signal and energy modulation for interacting electrons, which was also observed in all previous measured spectra, e.g., \cite{Peralta2013}. Unlike a strictly periodic DLA, where the modulation is symmetric due to uniform injection sampling of sinusoidal energy gain [Eq.  \ref{Eq:energyGain}], long multisegment APF DLAs result in an asymmetric signal according to the designed acceleration ramp, seen in both measured and simulated spectra.

APF DLAs ideally should have a prebuncher, which would introduce electrons with the correct injection phase (-$60^\circ$) with optimal laser amplitude, thus resulting in larger acceleration peak than seen in Fig. \ref{fig:MeasSim} \cite{tuprints22846}. In this experiment, there is no prebuncher, and the 830 fs FWHM electron pulse duration, compared to the laser's 330 fs, results in electrons sampling all laser amplitudes with an approximately uniform injection phase distribution. This complicates attributing any feature, such as the 2--3 keV deceleration shoulder common to all spectra, to a specific injection electron phase space or specific laser amplitude. The exception is the designed acceleration peak, which appears at or above the optimal field and with transverse emittance and injection phase within the accelerator's acceptance (i.e. dynamic aperture). See Supplemental Material for phase space evolution \cite{APF_DLA_Supplemental_2024}.

\begin{figure}[hbt!]
\includegraphics[width=8.5cm]{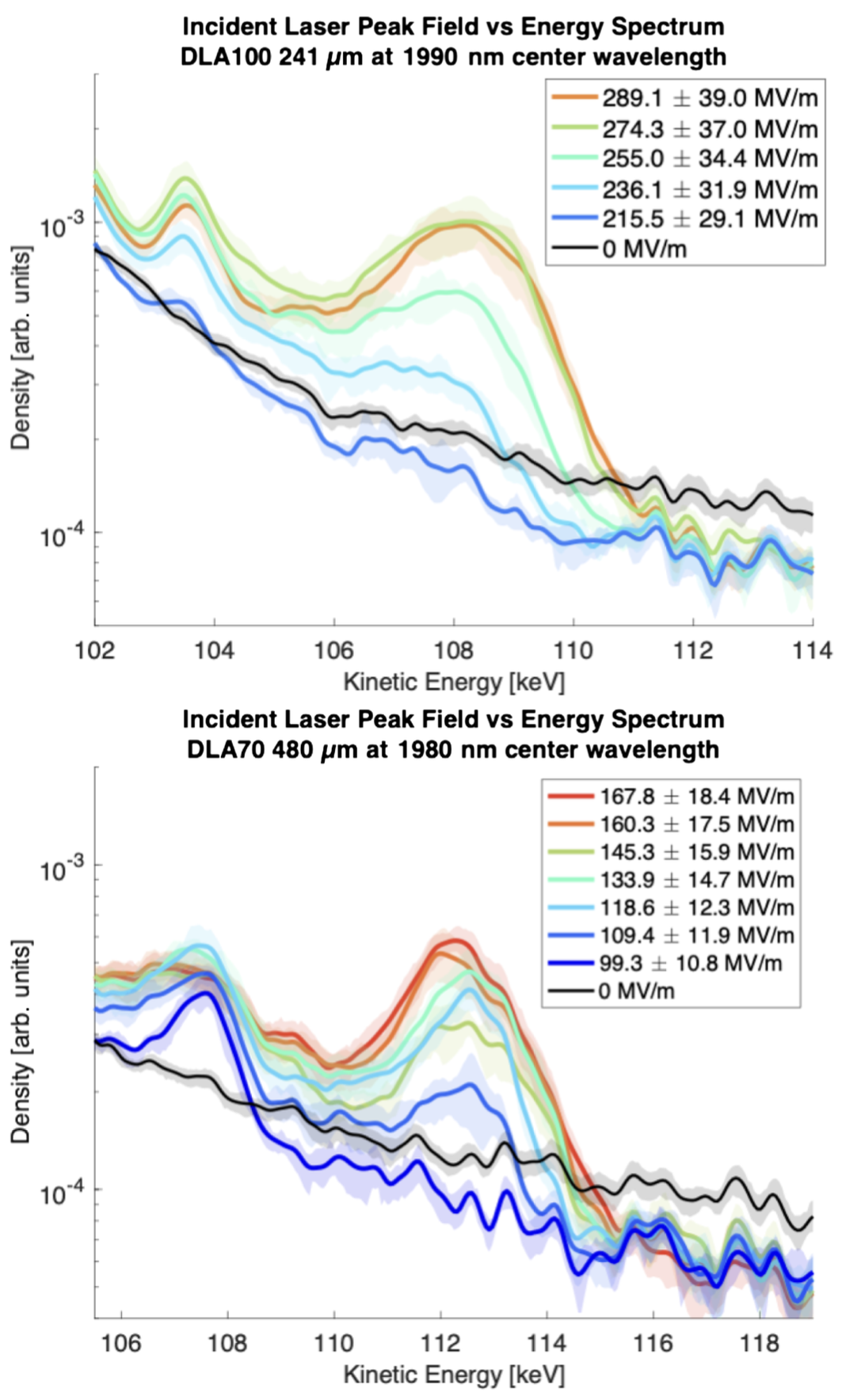}
\caption{Measured laser-on spectra for DLA70 and DLA100 with different peak electric fields. DLA70 and DLA100 operate optimally at peak electric field of 106 and 250 MV/m respectively. A semitransparent overlay on the measured spectra covers the 25-75\% quantile.}
\label{fig:TurnOn}
\end{figure} 

Figure \ref{fig:TurnOn} shows the acceleration spectra for DLA70 and DLA100 with increasing peak PFT electric field amplitude. DLA70 and DLA100 operate optimally with laser fields of 106 and 250 MV/m, respectively. With larger PFT drive field, more optical cycles of electrons interact with optimal or above fields, resulting in a larger capture peak. For DLA70, there is no visible capture peak at 99.3 MV/m and there is a visible capture peak at 109.4 MV/m, with peak increasing with larger PFT maximum fields. Similarly for DLA100,  there is no visible capture peak at 215.5 MV/m and there is a visible capture peak at 236.1 MV/m, also showing the same peak scaling, supporting APF operation. The proportion of arriving electrons that reach designed acceleration, or particle capture, is calculated by multiplying the simulated laser-off particle survival and the normalized capture peak from Fig. \ref{fig:MeasSim}. 
Captured current can be approximately obtained by multiplying input current ($\sim$ 8 fA) and capture rate. An upper limit estimate of $\sim$ 50-80 pmrad normalized $y$ transverse emittance is extracted from MCP transverse slices; cf. Supplement Material \cite{APF_DLA_Supplemental_2024}. This Letter makes no claims on x emittance.

DLA70 produced a larger accelerated population of electrons than DLA100 for similar lengths, and was easier to align. Furthermore, the DLA100 476 $\mu$m structure only worked at a longer wavelength of 1995 nm than designed 1980 nm, and required longer integration time to observe the signal.  It is unclear whether the pillar geometry, higher incident field requirements, or design differences resulted in this. Minor dual-drive phase differences generate sinh deflection forces, suppressed by highly transparent pillars in DLA70, but potentially resulting in significant sidewall losses in DLA100. 

In conclusion, we demonstrate a subrelativistic DLA architecture that enables extended energy gain over hundreds of optical periods, achieved through the capture and confinement of an electron bunch in the optical fields of a moving-bucket linear accelerator. These structures achieve coherent acceleration, i.e., uniform acceleration of a finite phase space volume, as opposed to a simple broadening of the energy spectrum peak. Good agreement between measured and designed energy gain was observed, with reduced measured versus simulated capture rate expected due to 3D defocusing effects intrinsic to finite pillar height DLA and field amplitude error from the pulse front tilted laser. Capture rate can be improved with electron macrobunching and mircobunching, such as the ones shown in ~\cite{PhysRevLett.127.164802,Niedermayer2018a,Niedermayer2021}, while longer interaction lengths should be possible with SOI-based 3D APF structures~\cite{3DAPF, PhysRevApplied.16.024022}. Similar results are shown in \cite{Chlouba2023}.

\begin{acknowledgments}
The authors wish to acknowledge the entire ACHIP collaboration for their support and guidance, as well as the staff from the Stanford Nanofabrication Facility (SNF) and Stanford Nanofabrication Shared Facilities (SNSF), supported by the National Science Foundation under Grant No. ECCS-2026822. T. Egenolf additionally acknowledge funding by the German Federal Ministry of Education and Research (Grant No. FKZ: 05K22RDC). This work is funded by the Gordon and Betty Moore Foundation (Grant No. GBMF4744).
\end{acknowledgments}





\pagebreak
\clearpage 
\widetext
\begin{center}

\textbf{\large Supplemental Materials: Subrelativistic Alternating Phase Focusing Dielectric Laser Accelerators}
\end{center}

\setcounter{equation}{0}
\setcounter{figure}{0}
\setcounter{table}{0}
\setcounter{page}{1}
\makeatletter
\renewcommand{\theequation}{S\arabic{equation}}
\renewcommand{\thefigure}{S\arabic{figure}}
\renewcommand{\bibnumfmt}[1]{[S#1]}
\renewcommand{\citenumfont}[1]{S#1}
\twocolumngrid

\maketitle
\section{Overview}

This supplement includes technical details for the APF DLA experiments and DLATrack6D simulations.

\maketitle

\section{Fabrication}

Using 100 keV electron beam lithography, both designs were patterned into ZEP520A on 5-10 $\Omega$cm B:Si silicon, with the longer structures consisting of multiple stitched 250 x 250 $\mu$m  write fields \cite{Miao:20}. A reactive ion etch (RIE) produced 2.7 $\mu$m tall pillars, gun sights, and apertures. A second photolithographic step followed by deep reactive ion etching (DRIE) produced 70 $\mu$m tall mesas. Each chip contained mesas 125 $\mu$m, 250 $\mu$m, 500 $\mu$m, 1 mm and 1.5 mm long, on which differing length DLA were included.

\section{Pulse Front Tilt}

\begin{figure}[h]
\includegraphics[width=7.3cm]{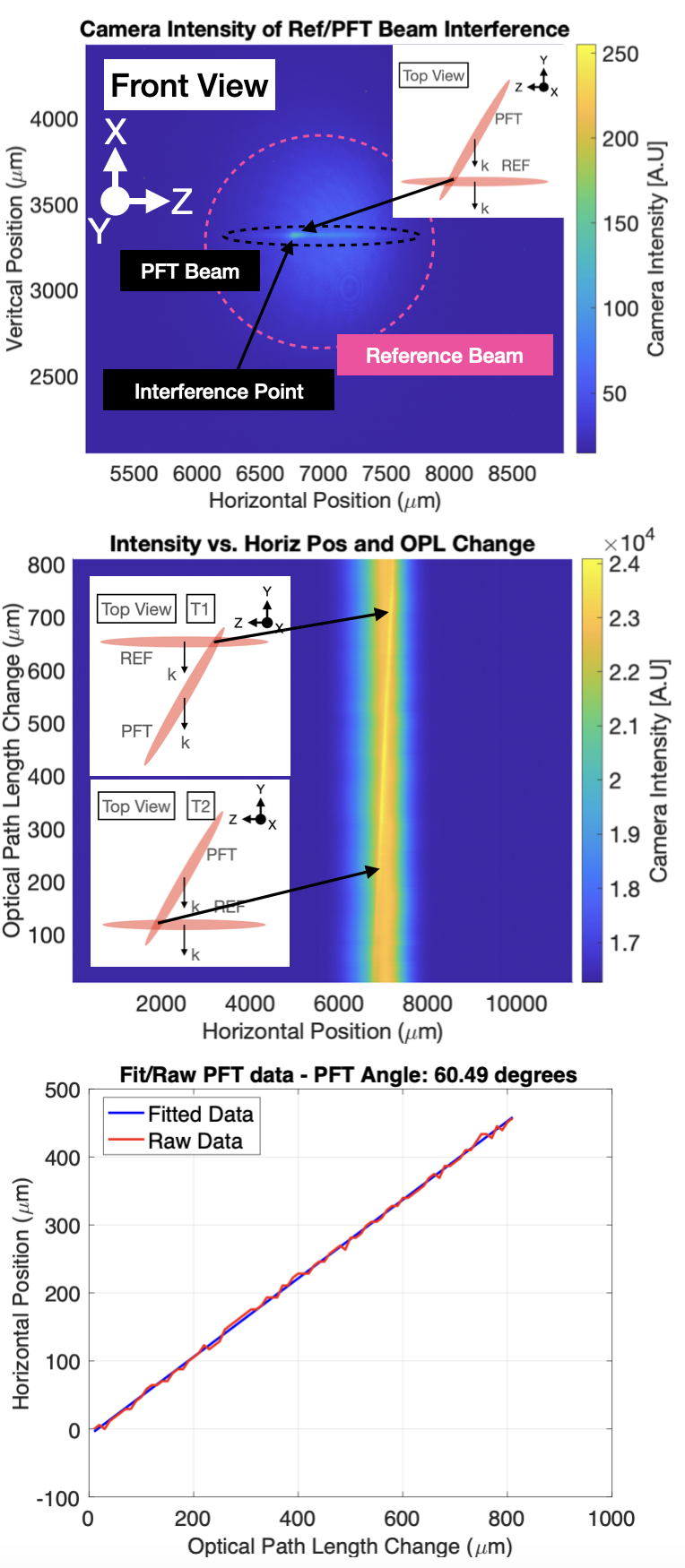}
\caption{Top: Camera image of reference and PFT beam at virtual interaction point. Middle: Camera intensity (vertical slice sums) versus horizontal position and optical path length change as the reference beam is swept across the PFT beam. Bottom: PFT angle fit. 
}
\label{fig:PFTChar}
\end{figure}

\begin{figure*}[t]
\includegraphics[width=17cm]{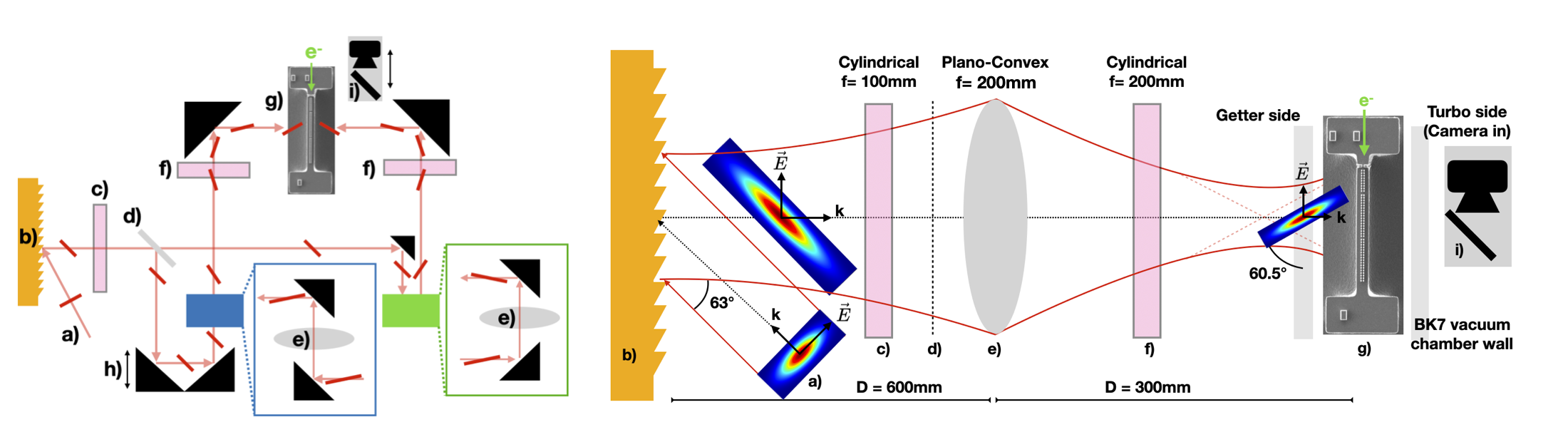}
\caption{Left: Optics overview. Right: PFT optics generation. Depending on the required final beam dimensions, different downstream expanders were used to produce different sized beams (a). Beam focus lands on 450 line/mm blazed grating (b) producing a PFT beam. The PFT beam travels through a single 100 mm cylindrical lens (c) before being split by a partially silvered mirror (d). Each beam then goes through a single 200 mm plano-convex lens (e) and 200 mm cylindrical lens (f) to strike a DLA (g). The getter side includes a delay line (h) to overlap beams in time and ensure equal arrival phase. The turbo side includes a drop in mirror and silicon camera (i) for beam dimension characterization.
}
\label{fig:PFTOpticsOnly}
\end{figure*}

Due to laser induced damage threshold (LIDT) scaling concerns and the chance our optical parametric amplifier (OPA) could not produce the necessary laser field amplitude with the available 2.3 -- 2.7 W power at 100 kHz repetition rate, a pulse front tilted (PFT) beam was used to illuminate our structures rather than a temporally stretched pulse \cite{Miao:20}\cite{2Soong2014}. To generate the first order diffraction (N = 1) for a 1980 nm beam on a 450 line/mm blazed grating (d = 1/450 mm), an incident angle $\theta_{bg}$ of 63$^{\circ}$ to normal was used \cite{2CESAR2018252,2Cesar:18,2Wei:17}. See equations below:
\begin{equation}
     \theta_{bg} = \text{sin}^{-1}(N\lambda_0/d),
    \label{Gratingangle}
\end{equation}
\begin{equation}
     \theta_{PFT} = \text{tan}^{-1}(1/\beta),
    \label{PFTangle}
\end{equation}
\begin{equation}
     m = \text{tan}(\theta_{PFT} )d/\lambda_0.
    \label{m}
\end{equation}

By illuminating the blazed grating at $63^\circ$, the first order (N = 1) diffraction lobe produces a PFT beam normal to the grating, which is re-imaged onto the DLA from both sides using a plano-convex lens to produce the correct PFT angle. This optical setup is designed for an angular magnification of $m=-2$ to produce a $60.56^\circ$ PFT beam at the DLA interaction point, chosen to group velocity match to a 108 keV beam, the average energy of electrons assuming a 24 keV energy gain. Fig. \ref{fig:PFTOpticsOnly} shows the optics used to produce $m=-2$.

Also included in the optical path was a 200 mm cylindrical lens and 100 mm cylindrical lens to reduce the vertical spot size while also not changing the horizontal PFT angle. These cylindrical lenses were added on each arm to reduce the vertical spot size at the DLA, resulting in less clipping from the mesa and higher available maximum electric field amplitude from the envelope. The position of these two cylindrical lenses were optimized to reduce the spot size on the beam profilometer (silicon camera) set at a virtual interaction point via a mirror pickoff on the turbo side of the accelerator. This camera was used to extract the beam profile. 

The $60.5 \pm 0.7^\circ$ angled beam at the DLA interaction point was measured by interfering the PFT beam and a non-titled replica reference beam on a delay stage at a virtual interaction point, as seen in Fig. \ref{fig:PFTChar}. A silicon camera (UX249) with pixel pitch 5.86 $\mu$m was put at the virtual interaction point. By sweeping the delay stage and taking intensity images, a horizontal location of max image intensity versus optical path length change dataset was gathered. A linear fit gave $60.5^\circ$ PFT angle and the fit residual gave $\pm 0.7^\circ$ standard deviation (Std).

\section{Electric Field Calculation and Error}

Most dual pillar DLA experiments included at least one section which was just an array of pillars and which could be illuminated independently of any other section. Testing this DLA would produce a spectrum with symmetric energy modulation, where the maximum energy gain ($\Delta W_{\text{max}}$) could be extracted from measuring the shoulder position on the acceleration spectrum. This maximum energy gain would result from electrons with zero interaction phase interacting with the maximum laser electric field ($E_{\text{max}}$): $\Delta W_{\text{max}}=\cos(0)E_{\text{max}} L e_{1n}$. With simulated structure factor $e_{1n}$ and structure length $L$ are already known, $E_{\text{max}}$ could be accurately calculated. Since no single dual pillar section could be isolated on our chip, we could not use this approach. Future experiments will include a simple long DLA test structure to better extract this value. Instead, we use a silicon camera at a virtual interaction point to extract a laser profile, which is used to calculate maximum field.

All laser beam measurements were also done with a UX249 silicon camera with 5.36 $\mu$m pixel pitch. Images were taken with 2 -- 3 second exposure sub-10db gain and with a ND1 filter to protect the camera from laser damage and a LP1350 filter to remove any residual 2nd harmonic noise. The camera was characterized to have a larger than 2nd order response to 1980 nm laser beam at these imaging conditions. Two images, a laser-off image and a laser-on image, were taken, with all dimension extraction using the image difference.

At these imaging conditions, the measured knife edge $x_{9010}$ laser intensity corresponded to within $\pm8\%$ of the $x_{9901}$ of the camera intensity. To calculate the maximum electric field amplitude of our PFT laser pulse, the spatial profile of the PFT was assumed to be spatially gaussian and temporally gaussian. With these assumptions we can calculate the peak electric field amplitude from one beam hitting one side of the DLA: 

\begin{equation}
E_{\text{peak}} = \left(\frac{8\sqrt{\ln 2}}{\pi^{3/2}}\frac{Z_0}{\tau_p f_{\text{rep}}}\frac{P_{\text{avg}}}{w_{0x}w_{0z}}\right)^{1/2}.
\label{Eq:MaxEField}
\end{equation}

Here, $\text{Z}_0$ is the impedance of free space (377 $\Omega$), $t_p$ is the FWHM of the temporal laser intensity profile (220 fs), $f_{rep}$ is the laser rep rate (100 kHz), $P_{avg}$ is the average laser power from one beam, and $w_{0z}$ and $w_{0y}$ are the beam radius ($I(x,z),\frac{I(w_{0x},0)}{I(0,0)}=\frac{I(0,w_{0z})}{I(0,0)}=1/e^2$) in $\hat{x}$ and $\hat{z}$ direction respectively. This equation was derived for a standard laser pulse, not a tilted PFT beam. However, it can be shown through simple trigonometry that this approach is still valid.

The average laser power is measured by multiplying a downstream pickoff power reading done before each experiment with the measured ratio from pickoff to DLA ($\sim$ $12.5\%$). To determine the beam waist sizes ($w_{0x}$ and $w_{0z}$), the laser beam is imaged, and $x_{9901CI}$ and $z_{9901CI}$ camera intensities are extracted. Given that $x_{9901CI}$ $\approx x_{9010}$, $w_{0x}$ and $w_{0z}$ are obtained as $x_{9901CI}/1.28$ and $z_{9901CI}/1.28$, respectively.  

We assume 30 fs standard deviation (Std) for temporal laser envelope and $10\%$ Std for $P_0$. We approximate the Std of $w_{0x}$ and $w_{0z}$ as $8\%$ of the measured width plus sum of one pixel length. We use error propagation to approximate the Std in $E_0$ by multiplying the Std of measured quantities with the partial derivatives of field with respect to a measured quantity. Equation \ref{Eq:EFieldError} shows this approach.

\begin{align}
\text{Std}_{E_0} &\approx \sqrt{\left(\frac{\partial E_0}{\partial P_0} \cdot \text{Std}_{P_0}\right)^2 + \left(\frac{\partial E_0}{\partial x_{9010}} \cdot \text{Std}_{x_{9010}}\right)^2} \nonumber \\
&\overline{\quad + \left(\frac{\partial E_0}{\partial z_{9010}} \cdot \text{Std}_{z_{9010}}\right)^2 + \left(\frac{\partial E_0}{\partial t_p} \cdot \text{Std}_{t_p}\right)^2}.
\label{Eq:EFieldError}
\end{align}

The resulting Std is approximately 10 -- 15\% of the measured electric field for different experiments. We calculated the $E_{peak}$ and Std of the electric field from the beam that hits the turbo side of the DLA. For simplicity, we assume the two beams illuminating our DLA, one from each side, have the same electric field for all DLATrack6D simulations. 

Table \ref{tab:DLAproperties} shows the different measured PFT dimensions for different DLA tests. The tests of DLA100 241 $\mu$m and 476 $\mu$m included a number of optical adjustments during the testing which do not accurately reflect the beam waist as measured pre-test. Instead, we assume the beam waist is the minimum of all measured images for the particular optical setup, and increase the Std of  $w_{0x}$ and $w_{0z}$ accordingly.

\begin{table}[ht]
\centering
\caption{PFT laser properties}
\label{tab:DLAproperties}
\begin{tabular}{|p{1.16cm}|p{1.5cm}|p{2cm}|p{2.5cm}|}
\hline
DLA Type & Length ($\mu$m) & Test wavelength (nm) & $z_{9010},x_{9010}$\\
\hline
DLA70 & 192 & 1980 & 1.64 mm, 53 $\mu$m \\
DLA70 & 480 & 1980 & 2.54 mm, 47 $\mu$m \\
DLA70 & 708 & 1980 & 2.81 mm, 53 $\mu$m \\
DLA100 & 241 & 1990 & 2.07 mm, 40 $\mu$m \\
DLA100 & 476 & 1995 & 2.07 mm, 40 $\mu$m \\
\hline

\end{tabular}
\end{table}

\section{Simulation Overview}

All simulations were accomplished with symplectic particle tracking software DLAtrack6D \cite{PhysRevAccelBeams.20.111302}. We use previously measured Glassbox beamline quantities to approximate the electron bunch in our  simulations: 830 fs FWHM temporal distribution, a transverse RMS focus of 0.41 $\mu$m, and a transverse RMS divergence of 390 $\mu$rad in $\hat{x}$ and $\hat{y}$ in the operating low charge regime \cite{10.1063/5.0086321}. For calculation simplicity, the electron beam is assumed to be temporally and spatially gaussian, with uniform transverse slice emittance i.e., same divergence and RMS focus for each temporal slice of the beam.

The PFT laser beam is also assumed to be gaussian temporally (220 fs FWHM intensity) \cite{10.1063/5.0086321}. For each experiment, the laser was imaged using the drop in mirror and pickoff power measured, which allowed $E_{max}$ to be extracted (see above). Our simulations use this $E_{max}$, and assume perfect alignment to the DLA. The beam is assumed to be uniform in $\hat{x}$ for the 2.7 $\mu$m tall pillars. Since the $z_{9010}$ intensity measurements are much larger than the DLA lengths, we can safely use a spatial top hat beam approximation for our DLA simulations.

\begin{figure}[h]
\includegraphics[width=8.5cm]{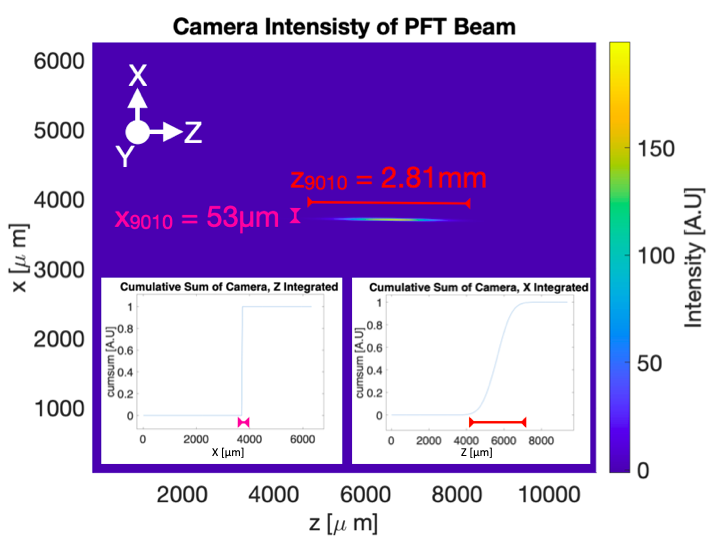}
\caption{Camera image of PFT beam used for DLA70 708 $\mu$m tests. Shown also are the extraction plots for light intensity $x_{9010}$ and $z_{9010}$.
}
\label{X9010}
\end{figure}

To create the acceleration spectra, the electron beam is first artificially split into bunches one optical cycle long in time i.e., $\Delta t = \frac{\lambda_0}{c} = 6.63$ fs. Since timing overlap tests are done on smaller test structures before each experiment, the two beams are modeled as arriving such that the middle of the electron packet interacts with the peak of the PFT beam. 

For each electron bunch, the average incident electric field amplitude $E_{inc}$ each bunch would experience is calculated. A random distribution of $1\cdot10^5$ bunch of 96 keV electrons with 100 pmrad slice emittance and uniform injection phase is then generated. These electrons are tracked through the structure, experiencing the $E_{inc}$ previously calculated. Due to the relatively small change in relativistic $\beta$ of the electron, the laser amplitude is assumed to be constant regardless of relativistic velocity-energy scaling.  All simulations apply a $5\%$ phase and amplitude uniform error to approximate wavefront error. 

Acceleration spectra of these bunches are formed by a sum of $\sigma_{\text{PSF}}=750$eV/($2\sqrt{2 \ln 2}$) gaussian's centered around each of the $N$ surviving electrons' particle energy $E_{i}$, which simulates the combined effect of the PSF of the sector magnetic spectrometer and the microchannel plate detector (MCP).
For each incident laser amplitude  $E_{inc}$, the spectra are constructed from $N$ surviving particles of the simulation as:
\begin{equation}
S(E, E_{\text{inc}}) = \sum_{i=1}^{N} \frac{1}{\sigma_{\text{PSF}}\sqrt{2\pi} } e^{-\frac{(E - E_i)^2}{2\sigma_{\text{PSF}}^2}},
\label{Eq:Spectrum}
\end{equation}
where $E$ is the kinetic energy.

\begin{figure}[h]
\includegraphics[width=8.5cm]{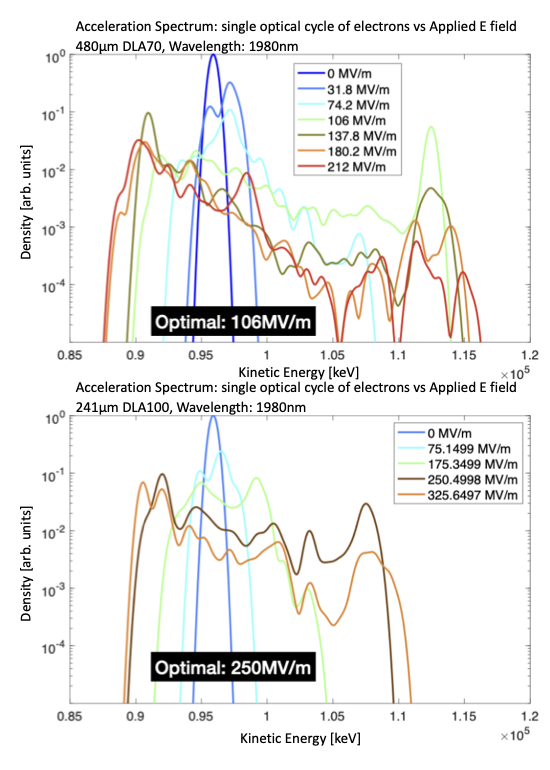}
\caption{Simulation of a single optical cycle of electrons through 241 $\mu$m DLA100 and 480 $\mu$m DLA70 respectively at different drive fields. 
}
\label{sectSim}
\end{figure}

Fig. \ref{sectSim} shows the $S(E, E_{\text{inc}})$ for different $E_{\text{inc}}$ for both DLA70 and DLA100. In these plots, $S(E, E_{\text{inc}})$ is normalized to the maximum of $S(E, 0)$. In both plots, all electrons in bunches that experience laser amplitude below the optimal laser amplitude are not captured, while some electrons in bunches that experience laser amplitude at or above the optimal laser amplitude are captured.

The full simulated spectrum, which corresponds to the experiment, is formed by a weighted sum of each bunch's spectrum (shown in Fig. \ref{sectSim}), where the weight is the percent of  electrons (PE) that each simulated bunch represents of the full beam, and the spectrum is simulated with the incident laser amplitude, $E_{inc}$, the respective part the electron bunch experiences: 

\begin{equation}
\text{PE}(t) = \frac{1}{2} \left( \text{erf}\left(\frac{t+\frac{\Delta t}{2}}{\sigma_{\text{ebeam}} \sqrt{2}}\right) - \text{erf}\left(\frac{t-\frac{\Delta t}{2}}{\sigma_{\text{ebeam}} \sqrt{2}}\right) \right),
\end{equation},

\begin{equation}
E_{inc}(t) = E_{max} e^{-\frac{t^2}{2  \sigma_{PFT}^2}} ,
\end{equation},

\begin{equation}
S_{\text{full}}(E, E_{\text{inc}}) = \sum_{n = -M}^{M} S(E, E_{\text{inc}}(n  \Delta t)) \cdot PE(n \Delta t),
\end{equation}

\begin{equation}
\sigma_{ebeam} = \frac{830 \, \text{fs}}{2 \sqrt{2  \ln(2)}}, \sigma_{PFT} = \frac{310 \, \text{fs}}{2\sqrt{2\ln(2)}}.
\end{equation}

We choose $M$ such that $\pm3 \sigma$ of the electron beam is simulated. In order to compare to the experiment, as done in the main paper, we normalize the laser-on spectrum to the peak of the laser-off spectrum as 
\begin{equation}
\begin{aligned}
S_{\text{full\_norm}}(E, E_{\text{inc}}) &= \frac{S_{\text{full}}(E, E_{\text{inc}})}{\max_{E}(S_{\text{full}}(E,0))} .
\end{aligned}
\label{normSpectra}
\end{equation}

\section{Capture Rate and Transverse Emittance Calculation}

Although the captured electron peak is easily distinguishable by its shape relative to the laser-off simulations and measurements and the peak maximum easily quantifiable, calculating the number of electrons captured requires recognizing where capture begins and ends in the energy spectrum. Due to the asymmetry most measured and simulated peaks show, we opt to logfit a gaussian to these peaks, and use the area under the fitted gaussian as a proxy for capture count.

\begin{figure}[h]
\includegraphics[width=8.5cm]{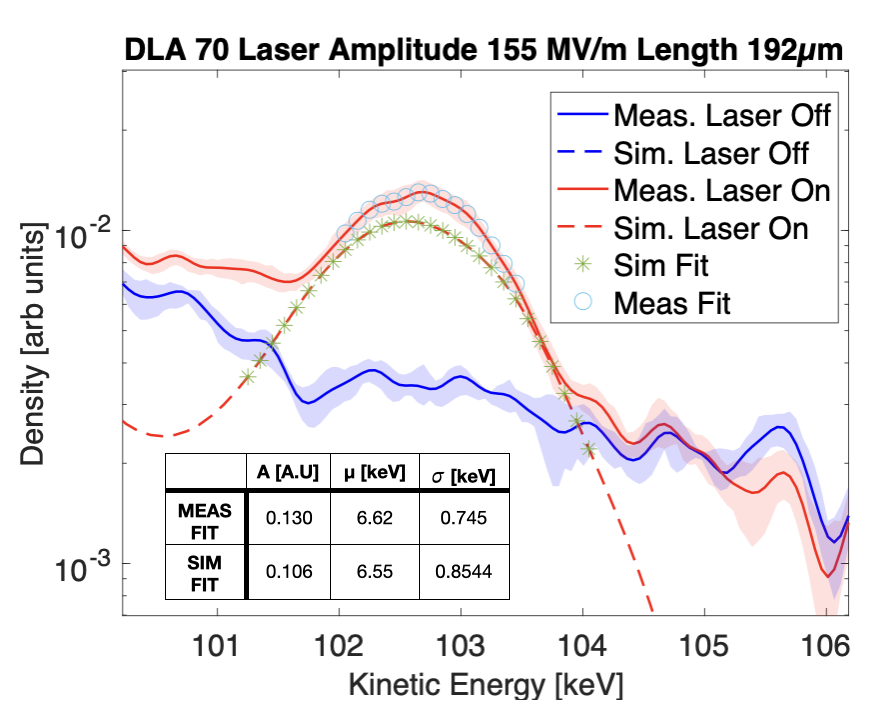}
\caption{Simulated and measured spectra DLA70 192 $\mu$m gaussian fit. The star and circle shows gaussian fit region with fit A, $\sigma$, and $\mu$ for simulated and measured fit respectively. 
}
\label{gaussFit}
\end{figure}

\begin{figure*}[t]
\includegraphics[width=16.5cm]{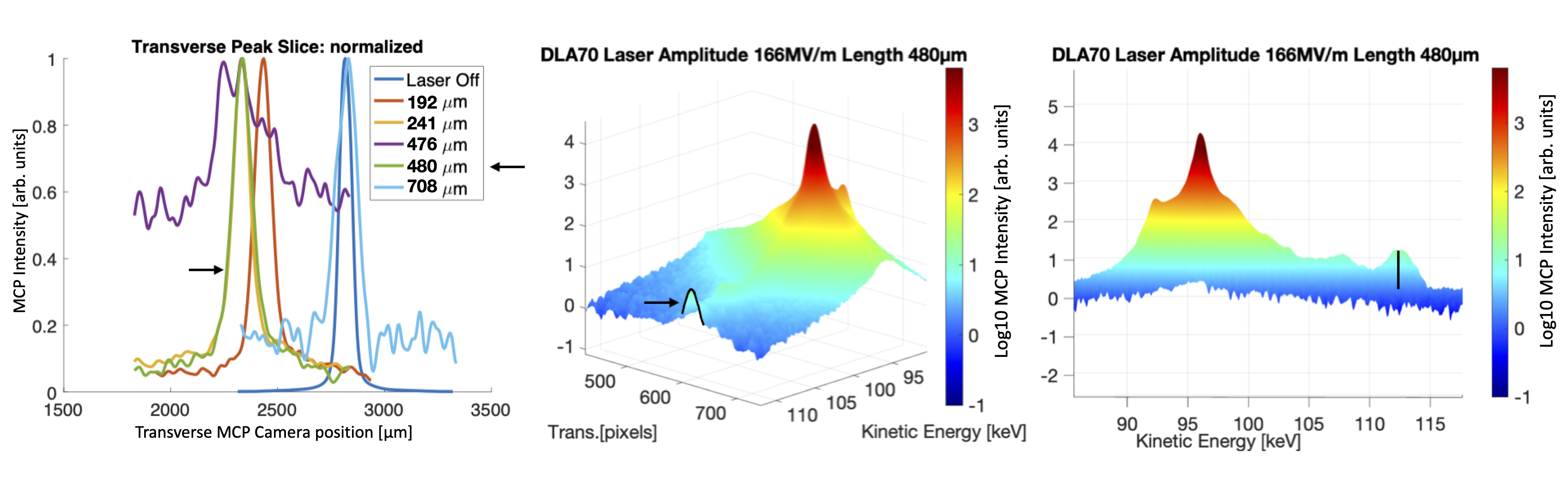}
\caption{Left: normalized MCP transverse slices of capture peaks and DLA70 192 $\mu$m laser-off reference versus camera location. Middle: surf plot of MCP 3D dataset [($\hat{x}$) kinetic energy, ($\hat{y}$) transverse pixel, ($\hat{z}$) Log10 MCP intensity] of DLA70 480 $\mu$m experiment. Black curve shows transverse slice of capture peak. Right: DLA70 480 $\mu$m surf plot obscuring transverse axis. Black line shows location of transverse capture slice.
}
\label{Transverse}
\end{figure*}

Fig. \ref{gaussFit} shows the fitted $\sigma_{\text{sim}}$ and $A_{\text{sim}}$ for the simulated DLA70 192 $\mu$m spectrum and the fitted $\sigma_{\text{meas}}$ and $A_{\text{meas}}$ for the measured DLA70 192 $\mu$m spectrum. Since the measured spectrum is formed by electrons that experienced the combined PSF of the magnetic spectrometer and MCP, and the simulated spectrum is convolved with a 750 eV FWHM gaussian to simulate the combined PSF of the magnetic spectrometer and MCP, the extracted $\sigma_{\text{sim}}$ and $\sigma_{\text{meas}}$ is not equivalent to the physical Std of energy. Instead, the simulated and measured captured energy Std can be calculated as:

\begin{equation}
\text{Std}_{\text{sim}} = \sqrt{(\sigma_{\text{sim}}^2 - (\frac{750eV}{2\sqrt{2\text{ln}(2)}})^2)}. 
\end{equation}
\begin{equation}
\text{Std}_{\text{meas}} = \sqrt{(\sigma_{\text{meas}}^2 - (\frac{750eV}{2\sqrt{2\text{ln}(2)}})^2)} 
\end{equation}

These values are multiplied by $2\sqrt{2\text{ln}(2)}$ to give the FWHM kinetic energy spread seen in Table 1 in the main text. The capture rate is calculated as the ratio of the area under the laser-on peak and area under the laser-off peak multiplied by the laser-off survival. The laser-off spectrum is normalized: $A_{\text{off}} = 1$ and $\sigma_{\text{off}} = 750\text{eV}/2\sqrt{2 \ln 2}$. Thus, given $\sigma_{\text{sim}}$, $A_{\text{sim}}$ and $\sigma_{\text{meas}}$, $A_{\text{meas}}$, the simulated and measured capture can be calculated as:

\begin{equation}
\text{Capture}_{\text{sim}}(L,A_{\text{sim}},\sigma_{\text{sim}}) = \text{Surv(L)}\frac{A_{\text{sim}}\sigma_{\text{sim}}}{1\frac{750eV}{2\sqrt{2 \ln 2}}}.
\end{equation}
\begin{equation}
\text{Capture}_{\text{meas}}(L,A_{\text{meas}},\sigma_{\text{meas}}) = \text{Surv(L)}\frac{A_{\text{meas}}\sigma_{\text{meas}}}{1\frac{750eV}{2\sqrt{2 \ln 2}}},
\end{equation}

Surv(L) is the simulated fraction of electrons that survive going through a laser-off DLA with length L. Since DLA70 has an aperture at the start of the channel with width 420 nm and DLA100 has an aperture at the start of the channel with width 400 nm, $\sim 60\%$ of electrons are lost at the entrance: $\text{Surv(0})_{DLA100} = 0.375$ and $\text{Surv(0})_{DLA70} = 0.391$. The simulated and measured capture current can be calculated using the capture rate and the electron source beam current $I_0$:

\begin{equation}
I_{\text{sim}} = I_0 \text{Capture}(L,A_{\text{sim}},\sigma_{\text{sim}}),
\end{equation}
\begin{equation}
I_{\text{meas}} = I_0 \text{Capture}(L,A_{\text{meas}},\sigma_{\text{meas}}).
\end{equation}

All experiments were done at $\sim$ 0.5 electrons per shot at 100 kHz repetition rate; approximately an $\sim$ 8 fA current, which gives on the order of aA (attoAmps) of captured electron current for our devices. 

Fig. \ref{Transverse} shows the transverse MCP profiles of the capture peak for laser-on spectra from Fig. 4 in the main text and the transverse profile of the 96 keV laser-off peak for DLA70 192 $\mu$m from Fig. 4 in the main text. We approximate the $y'$ divergence by fitting gaussian profiles to the laser-off and laser-on transverse slice profiles, and comparing the extracted sigma. The laser-off spectrum seen in Fig. \ref{Transverse} is produced by electrons traveling through a 420 nm wide 192 $\mu$m long DLA70 channel. With good alignment, this should only minimally filter the previously measured system beam divergence of RMS 390 $\mu$rad. Comparing the laser-on transverse sigmas to the laser-off DLA70 192 $\mu$m transverse sigma, which is assumed to correspond to RMS 390 $\mu$rad, gives RMS 617, 668, 848 $\mu$rad for DLA70 192, 480, and 708 $\mu$m respectively, and RMS 583 $\mu$rad for DLA100 241 $\mu$m respectively. The DLA100 476 $\mu$m measured peak is too close to the noise floor for an accurate divergence measurement. 

Without a physical knife edge scan, we can't extract $y$ focus, although worst case (full channel uniform distribution) gives 121 nm and 115 nm RMS focus $y$ for DLA70 and DLA100 respectively. Assuming $\langle y y' \rangle = 0$ correlation, an upper limit on normalized emittance can be calculated as: 

\begin{equation}
\epsilon_n = \gamma \beta \sqrt{\langle y^2 \rangle\langle {y'}^2 \rangle}.
\end{equation}

$\gamma$ is the Lorentz factor and $\beta$ the ratio between c and v for the electrons at the captured kinetic energy. This gives a normalized transverse $y$ emittance of 49, 49, 79 pmrad for DLA70 192, 480, and 708 $\mu$m respectively and 52 pmrad for 476 $\mu$m DLA100.

\begin{table}[ht]
\centering
\caption{Measured transverse $y,y'$ properties}
\label{tab:Trans}
\begin{tabular}{|p{1.16cm}|p{1.05cm}|p{1.15cm}|p{1.5cm}|p{2.4cm}|}
\hline
DLA Type & Length ($\mu$m) & $\text{RMS } y$ (nm) (sim)* & RMS $y'$ ($\mu$rad) (sim) & $\epsilon_n$ (pmrad) (sim w/wo correlation i.e., $\langle y y' \rangle = 0$)\\
\hline
DLA70 & 192 & 121(101) & 617 (1261) & 49 (67,85) \\
DLA70 & 480 & 121(82) & 668 (683) & 49 (33,39) \\
DLA70 & 708 & 121(66) & 848 (646) & 79 (23,31) \\
DLA100 & 241 & 115(96)& 583 (983) & 52 (64,65) \\ 
\hline
\end{tabular}
\textbf{*Note:} Not measured, worst case assumption. 
\end{table}

 Table \ref{Transverse} shows simulated and measured transverse $y,y'$ properties extracted from the same experiments described in Table 1 in the main text. Both the simulated and upper estimate measured normalized emittances are relatively close in value. Although the measured RMS $y'$ divergence for DLA70 increases with distance, the simulated divergence decreases, suggesting non-ideal performance at longer distances. For DLA100, the simulated $\langle y y' \rangle$ correlation is low, while DLA70 shows a more complex relation with correlation and $y$ emittance with distance.

\subsection{Capture Rate Error}

The capture calculation is done for both simulated and measured spectra, with simulated survival rate used for both. Using simulated survival rate was validated with transmission tests at low collection voltage: 47 $\mu$m DLA100 was measured to have $\sim40\%$ survival with simulated $\text{Surv(47)}_{DLA100} = 36.1 \%$ and 843 $\mu$m DLA100 was measured to have $\sim17\%$ survival with simulated $\text{Surv(47)}_{DLA100} = 16.3\%$.

The nonlinear response of the MCP driven at high collection voltage (1.7kV) results in the capture peak intensity depending on both the total global current hitting the MCP and local current near the peak and electrons that form the captured electron peak \cite{2leedle2016laser}. This can be seen in different background levels, where the laser-off background count is up to twice as large as the laser-on background. Without a detailed MCP characterization experiment with two electron beams, a current-tunable global beam to recreate the acceleration spectrum and accelerated beam with a fixed current at the design peak, it is difficult to determine whether the capture peak undergoes the same reduction as its background, or if there are other local scaling effects.

We model the MCP as linear, and capture area as correct, and include a potential $100\%$ error seen in the background in the error considerations for electron capture. 

We use error propagation to approximate the Std of capture by multiplying the Std of measured quantities with the partial derivatives of capture rate with respect to a measured quantity. Since $\text{Surv}$ is a simulated value rather than a measured value, we assume a potential $10\%$ Std for $\text{Surv}$ to account for misalignment. The Std for $A_{off}$ is simply the Std of the maximum values in the measured laser-off dataset (normalized to the mean maximum value), whereas the total Std for $A_{on}$ is modeled as the squared sum of the Std of $A_{on}$ and the potential $100\%$ error from MCP nonlinearity's: $\sqrt{std(A_{on})^2+(A_{on})^2}$. This error in $A_{on}$ dominates the calculation, seen in equation \ref{CaptureError}.

\begin{align}
\text{Std}_{\text{Capture}} &\approx \sqrt{\left(\frac{\partial \text{Capture}}{\partial \text{Surv}} \cdot \text{Std}_{\text{Surv}}\right)^2\nonumber} \\
&\overline{ + \left(\frac{\partial \text{Capture}}{\partial A_{off}} \cdot \text{Std}_{A_{off}}\right)^2} \nonumber \\
&\overline{\quad + \left(\frac{\partial \text{Capture}}{\partial A_{on}} \cdot \sqrt{\text{Std}_{A_{on}}^2+(A_{on})^2}\right)^2}.
\label{CaptureError}
\end{align}

 \section{Length capture effects}

 The DLA shown in the experiment are not infinitely scalable; with DLA70 708 $\mu$m the longest successfully tested DLA, although attempts continue with 842 $\mu$m and 1 mm DLA70. The effects that limit the length scaling also result in the reduced capture rate relative to simulated capture rate.

\begin{figure}[h]
\includegraphics[width=8.2cm]{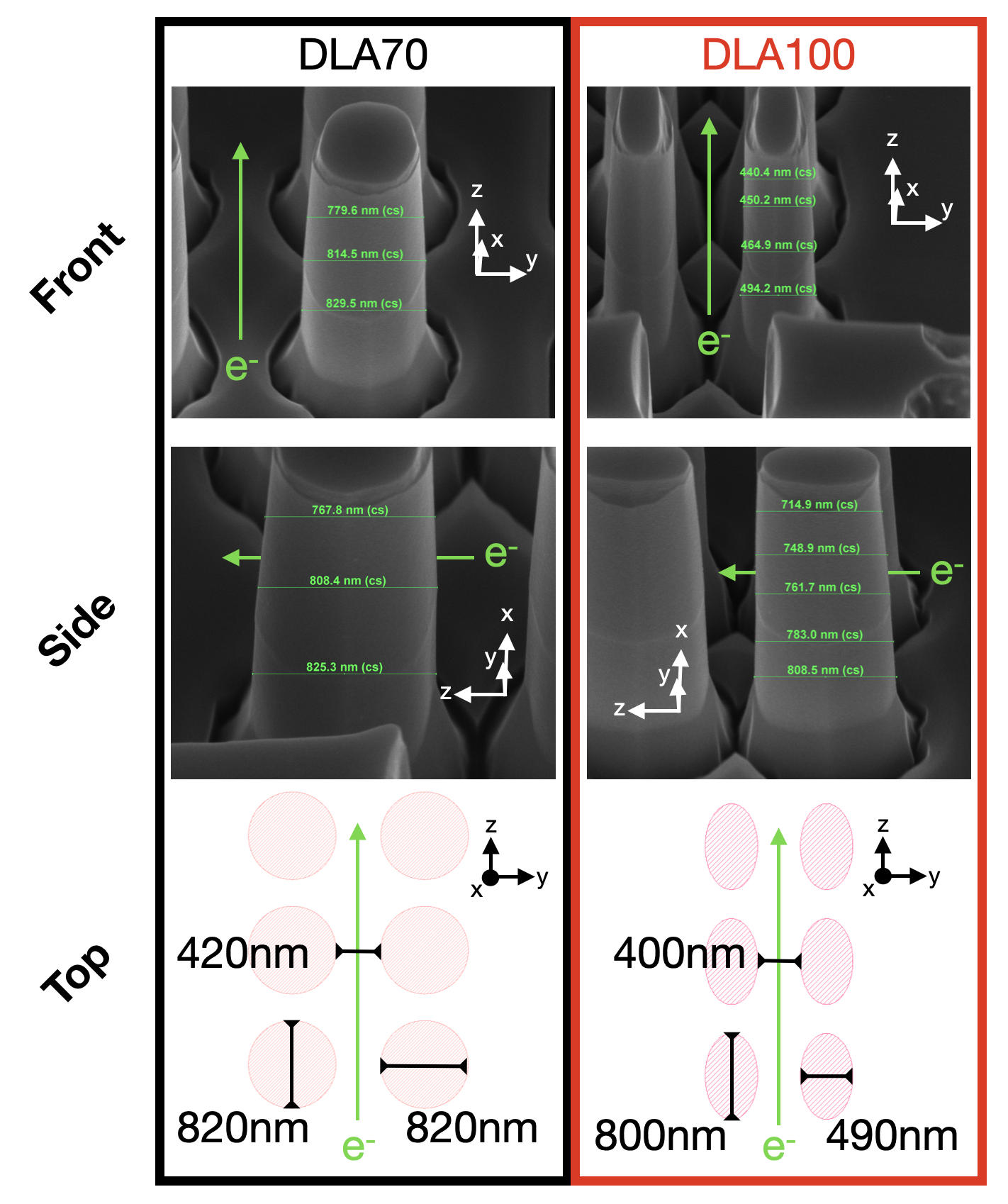}
\caption{Top and middle: Scanning electron micrographs of pillars at the entrance of DLA70 and DLA100. Bottom: DLA70 and DLA100 pillar design at entrance.  
}
\label{PillarSide}
\end{figure}
 Dual drive phase error from optical path length change due to optics drift has been previously estimated to be $\sim$ $\lambda_0/10$ with our optical setup \cite{2Black2022}. Any phase difference between drive lasers produces a deflecting sinh mode in the channel which result in unwanted transverse forces \cite{2PhysRevAccelBeams.23.114001}. Although zero mean random dual drive phase noise does produce some loss by driving electrons near the pillars into pillars and some capture reduction by pushing electrons out of the accelerating bucket, a non-zero mean dual drive phase results in sustained collective motion biased towards walls, resulting in potentially significant loss with increased length. This partially can explain DLA70 improved performance to DLA100, which has highly reflective pillar geometry that more easily supports sinh modes and a smaller channel width. Specifically, the simulated $r_1$ (the ratio of the two counter propagating evanescent accelerating modes) factor for DLA70 is 1.08 $\angle$ $-81^\circ$ while DLA100 has 0.04 $\angle$ $-177^\circ$. Efforts are underway to add random dual drive error and uneven dual drive amplitude to the DLATrack6D simulations for our structures for a more complete study. 

We suspect parasitic 3D effects are the most significant factor in producing capture rate loss and limiting millimeter scaling. Out of plane forces are expected due to the finite pillar height and the existence of the pillar mesa, which results in unwanted ${x}$ Lorentz forces within the channel. Although DLATrack6D does track the $x,x'$ of particles, out of plane forces are not included in the simulation. A rough fit of the $x$ dependence of the structure factors is possible to implement, however due to strong sensitivity on the exact 3D pillar shape it would not give  accurate results. 

Unexpected parasitic 3D effects include pillar geometry. Shown in Fig. \ref{PillarSide}, RIE etching of the electron beam resist resulted in a slight taper along the height of the pillar, resulting in additional $x$ dependent variation of the structure factor. The range in the semi-major and semi-minor radius of DLA100 (250 -- 215 nm) and (410 -- 350 nm) results in an initial structure factor range: 0.38-0.34 along the height of the pillar. For DLA70, the range in radius (420 -- 385 nm) results in an initial structure factor range: 0.49-0.63 along the height of the pillar. This trend, $\sim$ 10 -- 25 $\%$ structure factor variance on pillar height, continues through the entire structure. It is possible electrons are rejected in certain $x$ ranges, or more complex vertical electron motion through changing structure factors reduce capture rate. 

 \section{Electron motion and Spectra features}

DLA70 and DLA100 devices were designed to produce a capture peak from a correctly injected small emittance electron beam in an ideal structure at a fixed laser amplitude. In practical experiments with non-ideal conditions, these devices produce a rich acceleration spectrum that includes the intended capture peak alongside other distinctive features. To help the reader understand the origin of these features (i.e., the motion of the electrons in these non-ideal conditions that produce these spectra features), we include supplemental images (pngs) and videos (gifs) that display the phase space evolution of a single optical cycle of electrons, modeled with the slice emittance of the Glassbox beamline, with different incident laser amplitudes for DLA70 and DLA100 with different lengths.   

 \begin{figure}[h]
\includegraphics[width=8.2cm]{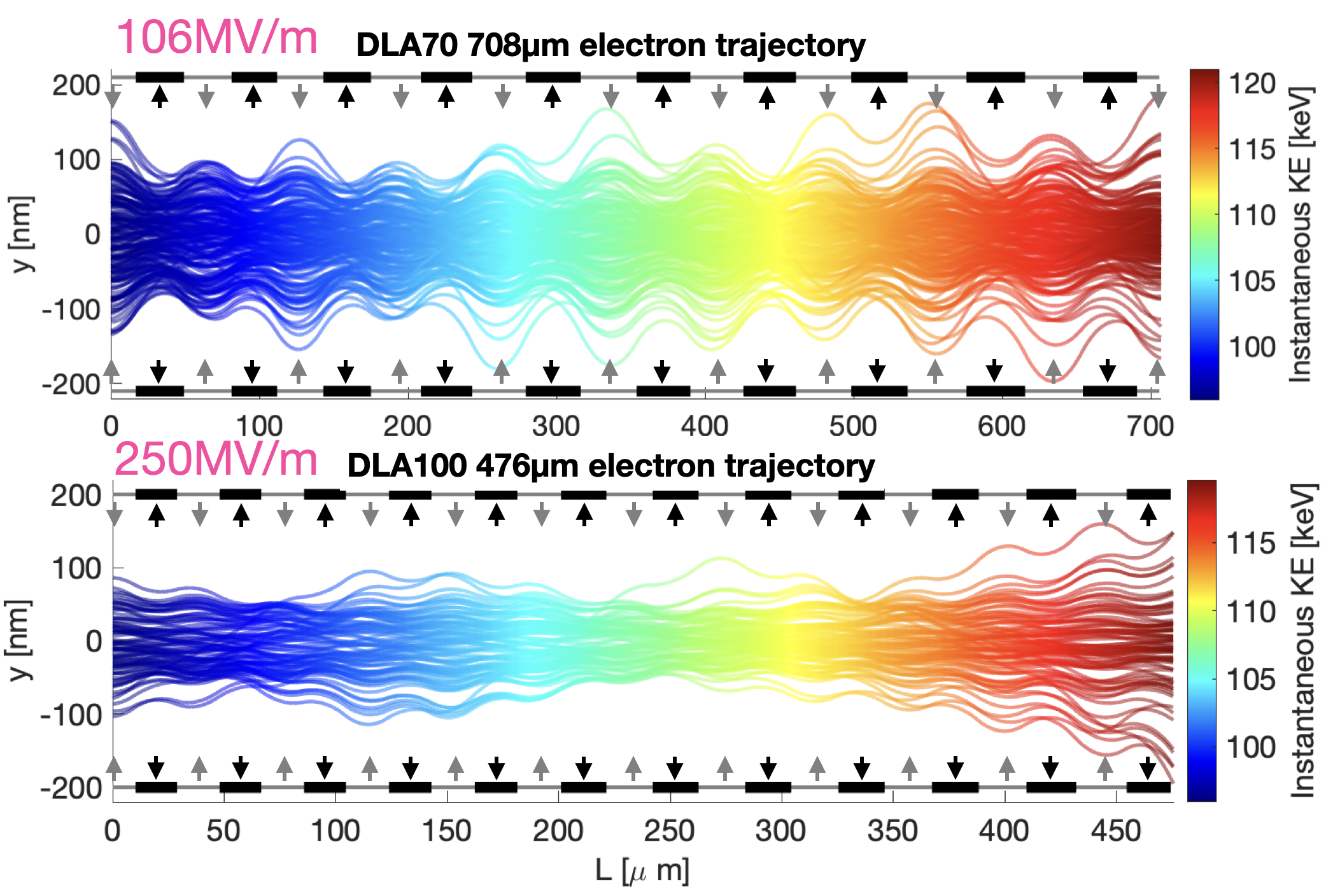}
\caption{Top: DLA70 708 $\mu$m simulated at optimal $E_0$ (106 MV/m) that reach within 500 eV of design energy gain $\Delta W = 24.39$ keV.  Bottom: DLA100 476 $\mu$m simulated at optimal $E_0$ (250 MV/m) that reach within 500 eV of design energy gain $\Delta W=23.51$ keV. The color corresponds to instantaneous energy. Simulation: DLATrack6D. E-Beam: N = 10,000, RMS focus = 410 nm, RMS div. = 390 $\mu$rad, Uniform injection phase $\phi_1 \sim \mathcal{U}(0, 2\pi)$. PFT Laser: wavelength = 1980 nm. Gray: transversely focusing longitudinally defocusing (TFLD). Black: transversely defocusing longitudinally focusing (TDLF).}
\label{ElectronMotion}
\end{figure}

\begin{figure*}[t]
\includegraphics[width=15cm]{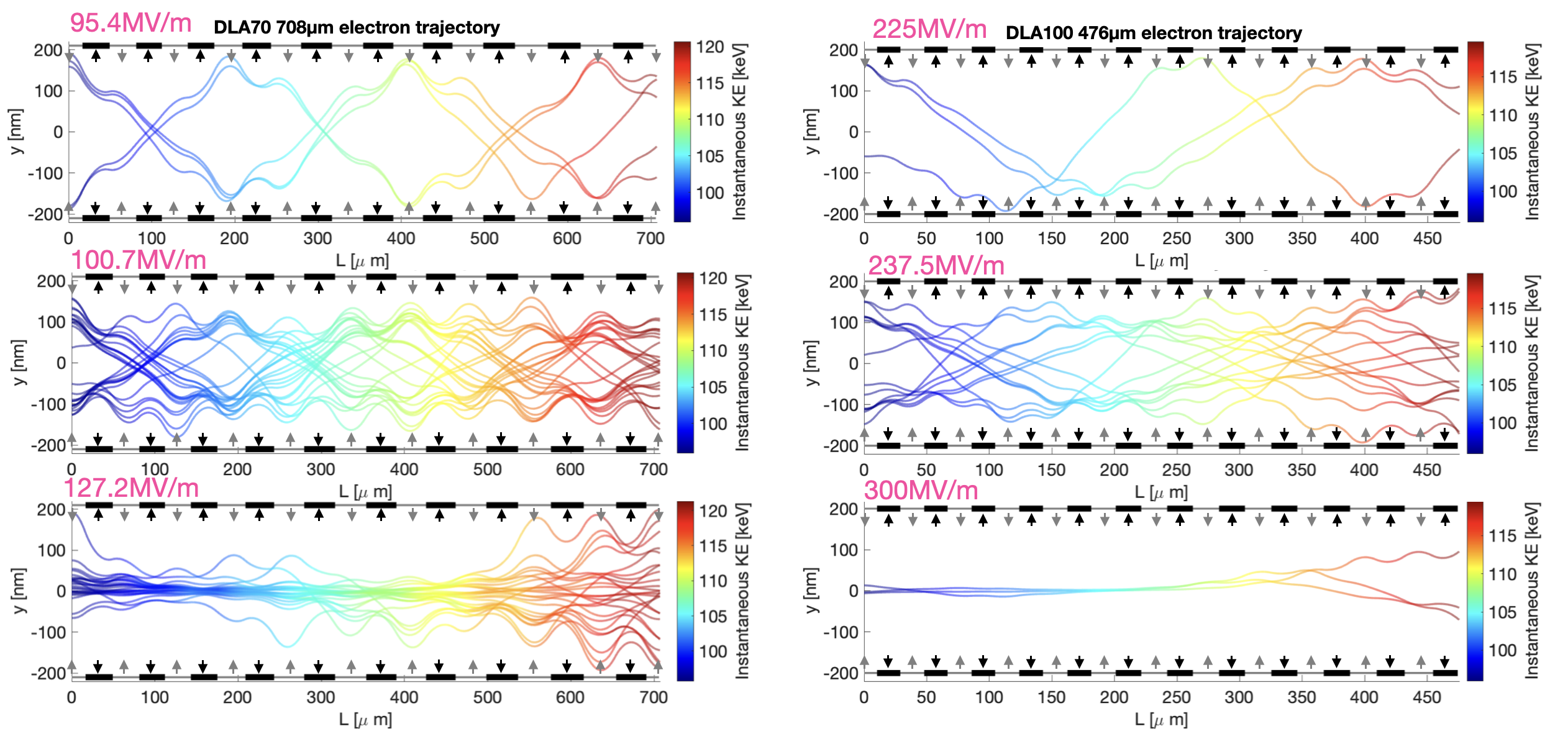}
\caption{Left: DLA70 708 $\mu$m simulated at x0.9, x0.95, and x1.2 optimal $E_0$ (106 MV/m) that reach within 500 eV of design energy gain $\Delta W = 24.39$ keV.  Right: DLA100 476 $\mu$m simulated at x0.9, x0.95 and x1.2 optimal $E_0$ (250 MV/m) that reach within 500 eV of design energy gain $\Delta W=23.51$ keV. The color corresponds to instantaneous energy. Same simulation parameters as Fig. \ref{ElectronMotion}.}
\label{ElectronMotionabovebelow}
\end{figure*}

The phase space trajectory requires tracking the 6D phase space ($\phi,W,y,y’,x,x’$) of N particles each experiencing unique laser field $E_0$. This is an 8D description (particle number,$\phi,W,y,y’,x,x’,E_0$). However, since we assume $\hat{x}$ invariance, we ignore $x$ and $x’$. We opt to display changes in phase space for different fixed laser amplitudes. 

Electrons in the correct phase space volume centered on $y=0$, $\phi_1 = -\pi/3$, $y'=0$, where $\phi_1$ is the injection phase, and experiencing equal to or greater than design laser amplitude ($E_0 > E_{\text{optimal}}$) will follow the acceleration ramp. Shown in Fig. \ref{ElectronMotion} is the transverse electron trajectory for DLA70 708 $\mu$m and DLA100 476 $\mu$m illuminated at optimal drive field (106 and 250 MV/m) with a single cycle of electrons with Glassbox slice emittance injected that reach the target energy. The transverse electron trajectories clearly show the APF nature of these devices for our Glassbox system electrons. 

Most electrons regardless of the laser amplitude are not in an accelerating bucket (i.e., a phase space volume centered on $y=0$, $\phi_1 = -\pi/3$, $y'=0$). These electrons do not follow the design energy gain ramp, resulting in the drift sections producing incorrect $\phi$ jumps. Thus, these electrons can be modeled as experiencing a pseudo-random interaction phase, resulting in some semi-symmetric energy broadening around the injection energy (96 keV), as seen in the low laser amplitude phase space plots in the supplemental images (pngs). 

Bunches with a laser amplitude significantly below the optimal laser amplitude ($E_{\text{optimal}}$) do not reach the design energy gain, thus producing relatively small semi-symmetric energy broadening. However, at $E_0$ closer to $E_{\text{optimal}}$ some electrons do reach high energies, with a spectra biased towards energy gain. Due to the $\text{Cosh}(\Gamma_y y)$ dependence on $F_z$ for the Cosh mode, electrons on the outskirts of the channel experience a higher $F_z$ than electrons in the center. For DLA70 with gap = 420 nm, $\text{Cosh}(\Gamma_{1y}(\beta=0.54)|210e^{-9}|)=1.59$ and $\text{Cosh}(\Gamma_{1y}(\beta=0.63)|210e^{-9}|)=1.356$, and for DLA100 with gap = 400 nm, $\text{Cosh}(\Gamma_{1y}(\beta=0.54)|200e^{-9}|)=1.53$ and $\text{Cosh}(\Gamma_{1y}(\beta=0.63)|200e^{-9}|)=1.33$.

With this 33-59\% increase in the acceleration gradient on the outskirts of the channel versus the center, electrons can compensate for lower $E_0$ to  follow the acceleration ramp, which was designed for electrons in the center of the channel.  Transverse forces can push the electrons into the pillars (loss) and can push the electrons back into the center of the channel, where they subsequently fall off the acceleration ramp. Electrons must spend most of their trajectory far from center (swings from side to side of the channel or stays on one side) to continue following the acceleration ramp. For $E_0$ below $E_{\text{optimal}}$, almost no electrons reach the target energy gain but some do reach large energies, helping to contribute to the acceleration arm and subpeaks. For $E_0$ above $E_{\text{optimal}}$, the opposite must occur; to follow the acceleration ramp electrons must move close to the center where $F_z$ is lowest and longitudinal focusing can best compensate. Both of these effects, electrons traveling near the edges of channel to compensate for low $E_0$ and moving to the center channel to compensate for high $E_0$, are seen in Fig. \ref{ElectronMotionabovebelow} for DLA70 and DLA100. 

With injection phase far from the optimal injection phase $\phi_1 = -\pi/3$, electrons quickly experience the pseudo-random interaction phase jumps, producing a semi-symmetric peak broadening. Due to the chirping of the underlying structure, the spectra can be biased towards deceleration. As seen in the supplemental images (png), for DLA100, electrons injected with $\phi_1 \approx \pi$ produce a deceleration signal and for DLA70, electrons injected with $\phi_1 \approx 2\pi/3$ produce a deceleration signal. Outside these points in phase space, the motion is more complicated.

\begin{figure}[h]
\includegraphics[width=8.5cm]{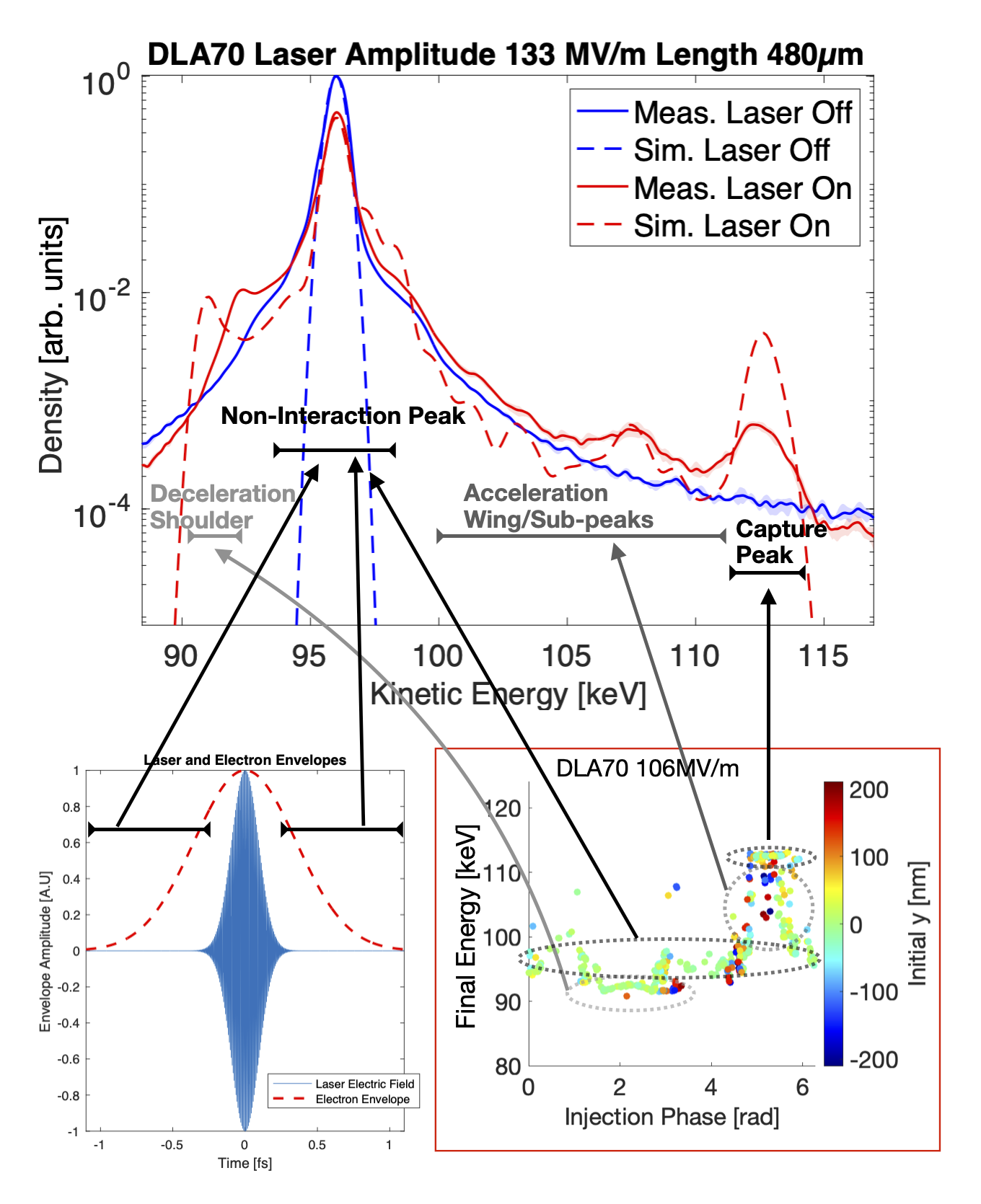}
\caption{Spectra feature origins for DLA70 480 $\mu$m. Top: Simulated and measured DLA70 480 $\mu$m reproduced from main text. Bottom left: Laser and electron envelopes. Bottom right: Single optical cycle of electrons simulated at optimal $E_0$ (106 MV/m) through DLA70 480 $\mu$m.  }
\label{fig:DLA70480explained}
\end{figure}

The majority of electrons forming the subpeaks occur due to large ``spills" from the accelerating ``bucket." As seen in each supplemental image's (png's) second plot for each laser amplitude with $E_0 \geq E_{\text{optimal}}$ (for instance the bottom right plot of Fig. \ref{fig:DLA70480explained}), where the plot shows injection phase versus final energy with initial $y$ color coded, a large number of electrons centered at injection phase $\phi_1 =-\pi/3$ form a shelf of electrons at the final design energy gain ($\Delta W$), with electrons underneath this shelf spanning a range of energies. The electrons at the design energy have coherently accelerated and followed the design energy ramp.  Electrons under the coherently accelerated electron shelf have traveled with these electrons over many APF periods before falling behind in energy. This shelf and lost electrons can be seen in Fig. \ref{fig:DLA70480explained} and on many supplemental images (pngs). These spills normally occur at the transition from TDLF to TFLD. These electrons are on the edge of the phase space acceptance window volume, and as the acceptance window changes, they fall out of the bucket. This is difficult to show graphically, but it can be easily shown with the animated videos (gifs) of the phase space evolution.

Our design analysis only considered the linear and constant component of the synchronous mode, not the higher order terms: $\mathcal{O}(\Delta \phi, \Delta y)$, which cause the actual, not ideal, acceptance phase space volume to change. An electron might be on the edge of the acceptance window, accelerating with the synchronous reference particle, only for the window to change. Once outside the window, this electron no longer experiences the correct acceleration ramp, and thus the synchronous electrons move away (gain energy) from the electron that now begins to experience pseudo-random interaction phase jumps (producing close to zero energy gain). 

Capture peak: Electrons with phase space volume centered on $y=0$, $\phi_1 = -\pi/3$, $y'=0$ and with $E_0 \geq E_{\text{optimal}}$ are captured and accelerated to high energy. Some electrons traveling mostly on the edge of the channel at low $E_0$ achieve high energy gain and also contribute to the capture peak. 

Acceleration wing and subpeaks: For $E_0 \geq E_{\text{design}}$, electrons on the edge of the phase-space volume of the acceleration bunch periodically fall out of the acceleration bucket. These electrons begin experiencing pseudo-random phase jumps producing slight peak broadening around the energy they fell out at. This contributes to the subpeaks on the acceleration spectra. For $E_0 < E_{\text{design}}$, some electrons traveling at the edge of the channel (large $|y|$) experience large energy gains before moving into the center of the channel and falling off the acceleration ramp and contributing to the acceleration wing and subpeaks. 

Deceleration wing and shoulder: Drift sections for non-captured electrons can be modeled as producing pseudo-random interaction phase jumps. Certain volumes, such as a large bunch injected $\pi$ to $2\pi/3$ behind design injection phase $\phi_1 = -\pi/3$, coupled with structure chirping produced sustained net deceleration. 

Non-interaction peak: For our system, the electron FWHM $\tau_e$ $>$ laser FWHM $\tau_{\lambda}$ (see lower left plot in Fig. \ref{fig:DLA70480explained}); a large percentage of electrons experience a small or zero $E_0$ and thus produce a small or zero $\Delta W$ and thus a large non-interaction peak at 96 keV. Psuedo-random interaction phase jumps for incorrectly injected electrons ($\phi_1 \neq -\pi/3$) for higher $E_0$ also produce low $\Delta W$, as seen in Fig. \ref{fig:DLA70480explained}.

This is only an overview of the general trends of electron motion from the simulations. Fig. \ref{fig:DLA70480explained} shows the contribution of only one optical cycle of electrons simulated at optimal $E_0$ (106 MV/m) to the different features of the full spectra. See the supplemental images (pngs) and videos (gifs) for finer details, where the same analysis shown in Fig. \ref{fig:DLA70480explained} can be applied to each plot. More information on the challenges of simulating and describing DLA electron motion is discussed in \cite{2Niedermayer:ICAP2018-MOPLG01}. 


\end{document}